\documentclass[a4paper,12pt]{article}
 \pdfoutput=1
\usepackage{amsmath,amssymb}
\usepackage{hyperref}
\usepackage{graphicx}
\usepackage{mathrsfs}
\usepackage{float}
\usepackage{ulem}
\hypersetup{
    colorlinks,%
    citecolor=blue,%
    filecolor=blue,%
    linkcolor=blue,%
   urlcolor=blue,
   linktoc=page
}
\usepackage{chngcntr}
\counterwithin{equation}{section}

\newcommand{\be}{\begin{equation}}
\newcommand{\ee}{\end{equation}}
\newcommand{\beq}{\begin{equation}}
\newcommand{\eeq}{\end{equation}}

\newcommand{\bea}{\begin{equation}\begin{aligned}}
\newcommand{\eea}{\end{aligned}\end{equation}}
\newcommand{\ba}{\begin{align}}
\newcommand{\ea}{\end{align}}

\begin{document}

\begin{titlepage}
%

\vspace{.4cm}
\begin{center}
\noindent{\Large \textbf{Echoes of chaos from string theory black holes}} 
\\
\vspace{1.5cm}
Vijay Balasubramanian$^{a,b}$, Ben Craps$^b$, Bart{\l}omiej Czech$^c$ and G\'abor S\'arosi$^{a,b}$

 \vspace{1.5cm}
  {\it
  $^{a}$David Rittenhouse Laboratory, University of Pennsylvania,\\
  Philadelphia, PA 19104, USA\\
\vspace{0.2cm}
 }
\vspace{.5cm}
 {\it
 $^{b}$Theoretische Natuurkunde, Vrije Universiteit Brussel (VUB), and \\ International Solvay Institutes, Pleinlaan 2, B-1050 Brussels, Belgium\\
\vspace{0.2cm}
 }
 \vspace{.5cm}
  {\it
 $^{c}$Institute for Advanced Study, Princeton, NJ 08540, USA\\
\vspace{0.2cm}
 }
\vskip 15mm
{\small\noindent  {\tt vijay@physics.upenn.edu, Ben.Craps@vub.ac.be, czech@ias.edu, gsarosi@vub.ac.be}}
\end{center}

\vfill
\begin{abstract}
The strongly coupled D1-D5 conformal field theory is a microscopic model of black holes which is expected to have chaotic dynamics.   Here, we study the weak coupling limit of the theory where it  is integrable rather than chaotic.   In this limit, the operators creating microstates of the lowest mass black hole are known exactly.  We consider  the time-ordered two-point function of light probes  in these microstates, normalized by the same two-point function in vacuum. These correlators display a universal early-time decay followed by late-time sporadic behavior.  
To find a prescription for temporal coarse-graining of these late fluctuations
we appeal to random matrix theory, where we show that a progressive time-average smooths the spectral form factor (a proxy for the 2-point function) in a typical draw of a random matrix.  This coarse-grained quantity reproduces the matrix ensemble average to a good approximation. Employing this coarse-graining in the D1-D5 system, we find that the early-time decay is followed by a dip, a ramp and a plateau, in remarkable qualitative agreement with recent studies of the Sachdev-Ye-Kitaev (SYK) model. We study the timescales involved, comment on similarities and differences between our integrable model and the chaotic SYK model, and suggest ways to extend our results away from the integrable limit.   
\end{abstract}

\end{titlepage}

\tableofcontents


\section{Introduction}

Black holes are the most entropic objects in the universe.   Their entropy $S_{BH} = {A_{BH} \over 4 G_N \hbar}$ is proportional to the horizon area and implies that the energy spectrum of microstates has a miniscule gap ($\delta E \sim e^{-A_{BH}/4G_N \hbar}$), which becomes infinitesimal in the classical $\hbar \to 0$ limit.  Various lines of evidence also suggest that the dynamics of the Hamiltonian acting on these microstates is chaotic \cite{Sekino:2008he,Shenker:2013pqa,Polchinski:2015cea}, implying that the spectrum of excitations must be irregular \cite{bohigas1984characterization}.  
Around a typical state of such a bounded system, general arguments from quantum mechanics suggest that the gapped, irregular spectrum will lead to temporal correlations showing  a  universal initial decay which gives way at very late times to rapid, small fluctuations whose precise structure is determined by the actual microstate. 

 Random matrix theory (RMT), where the Hamiltonian is drawn from a fixed ensemble, has been proposed as a universal description of this sort of behavior.  In this theory it has been shown that the ensemble average of the spectral form factor, a proxy for the two-point function related to the `easy' version of the information paradox \cite{Maldacena:2001kr,Papadodimas:2015xma,Papadodimas:2015jra}, exhibits a characteristic  initial decay, followed by an increasing ramp, and then a plateau.        Recently, it was shown that the Sachdev-Ye-Kitaev (SYK) model \cite{Sachdev:1992fk, Kitaev,Polchinski:2016xgd}, which is a strongly coupled model of quenched disorder inspired by black hole physics\footnote{This model exhibits a tractably broken Virasoro symmetry \cite{Maldacena:2016hyu} and its effective action takes a form which also arises in two-dimensional dilaton gravity \cite{Jensen:2016pah,Maldacena:2016upp,Engelsoy:2016xyb,Cvetic:2016eiv}.}, also displays the decay, ramp and plateau phenomena \cite{Cotler:2016fpe}.\footnote{In fact, the level statistics in SYK is well approximated by RMT around the mean level spacing \cite{Garcia-Garcia:2016mno,Cotler:2016fpe}.}  Like random matrix theory, the SYK model contains an average over Hamiltonians -- the coupling of the theory is drawn from a distribution and the smooth ramp and plateau arise after averaging over this ensemble.     We would like to understand whether this behavior occurs generally in black hole physics. 

  String theory contains many examples of black holes whose microscopic descriptions are well understood.  The simplest setting is Type IIB string theory compactified on a torus with five asymptotically flat dimensions.   This theory contains charged black holes whose extremal limit still has a large entropy.  The low-energy excitations of such a black hole are described by a two-dimensional conformal field theory~\cite{Strominger:1996sh}, the ``D1-D5 CFT'', which is a marginal deformation of a sigma model on the symmetric product target space $(T^4)^N/S_N$. Here $S_N$ is the permutation group acting on $N$ copies of $T^4$.  The marginal deformation parameter 
acts as the coupling in the theory.  When it is large the theory is expected to be chaotic as it describes a macroscopic black hole.   When it goes to zero, the theory approaches the symmetric product limit where it is integrable.  

These five dimensional black holes can be reduced through a sequence of near-horizon, low-energy limits to extremal black holes in three and two dimensional Anti-de Sitter space.  Indeed, through these limits, the D1-D5 CFT is known to be exactly dual to type IIB string theory on AdS$_3 \times S^3 \times T^4$ \cite{Maldacena:1997re,Maldacena:1998bw}.    The SYK model was inspired by AdS$_2$ black holes, and hence it may be that the  D1-D5 CFT at finite coupling has a reduction to an SYK-like model \cite{Maldacena:2016upp,Cvetic:2016eiv}.   Here, we instead study the weak coupling limit of the theory.   In this limit, the theory is integrable rather than chaotic but, remarkably, we show that many of the qualitative features of the chaotic RMT and SYK dynamics are already present.

Specifically, we consider dynamics around Ramond ground states of the D1-D5 theory which are typical microstates of the lightest black hole.   These black holes have a large microscopic entropy, although it is not large enough to produce a classical black hole horizon.  The temporal correlation function of graviton operators in these states shows an initial universal decay followed by sporadic fluctuations \cite{Balasubramanian:2005qu}.    A similar structure occurs in observables computed with a {\it single} draw of a Hamiltonian in a random matrix theory.  We argue, and numerically demonstrate, that  the ensemble average in RMT can be mimicked by a progressive time-average  in a single draw from the theory, over windows that scale proportionally to time.    Applying this progressive time average to correlators in a typical ground state of the D1-D5 theory reveals an initial decay, followed by a long ramp and a plateau, qualitatively resembling both the RMT and SYK theories.   The initial decay exactly reproduces the expected results in a black hole background \cite{Balasubramanian:2005qu}.   We present analytic calculations of the plateau height and the shape of the ramp, and comment on the reasons for the quantitative differences between our results and those in fully chaotic theories like RMT and SYK.  An interesting challenge for the future is to perturbatively turn on the marginal deformation that takes the integrable limit of the D1-D5 theory into a chaotic regime.

The paper is organized as follows. In section \ref{sec:d1d5corr} we give a short review of the D1-D5 system at the orbifold point and the two point functions in the Ramond ground states based on \cite{Balasubramanian:2005qu}. In section \ref{sec:rmt} we introduce progressive time averaging in the context of random matrix theory and present evidence that it is capable of capturing both qualitative and quantitative features of the ensemble average. In section \ref{sec:d1d5latetime} we apply this time average to the D1-D5 graviton two point function and present the main results of the paper, including analytic estimates for the ramp and the plateau. In section \ref{sec:disc} we conclude the paper. We include two appendices with some additional details of the discussion in section \ref{sec:d1d5latetime}.

\section{Correlators in the D1-D5 CFT at the orbifold point}
\label{sec:d1d5corr}

We will consider Type IIB string theory compactified to five dimension on $S^1 \times T^4$ with $N_1$ D1-branes wrapped on the $S^1$ and $N_5$ D5-branes wrapped on the entire compact space.  At low energies, the effective theory describing the dynamics of excitations is a certain marginal deformation of an ${\cal N} = (4,4)$ supersymmetric sigma model on the symmetric product target space ${\cal M}_0 = (T^4)^N/S_N$, where $N=N_1 N_5$ and $S_N$ is the permutation group acting on the $N$ copies of $T^4$ \cite{Strominger:1996sh}.  The sigma model on ${\cal M}_0$ describes $N$ ``strands'' of string propagating on $T^4$.  While this is a free orbifold theory, it has an interesting spectrum and correlation functions, as we will see. The marginal deformation corresponds to turning on an interaction that allows splitting and joining of strings.

Taking an appropriate limit isolates the part of the spacetime that is exactly described by the CFT.   In this low-energy limit we say that the D1-D5 CFT is holographically dual to Type IIB string theory on ${\rm AdS}_3 \times S^3 \times T^4$.     To have a large, weakly coupled AdS$_3$ space, $N$ must be large and, in addition, the CFT must be strongly coupled, i.e.\ deformed far from the orbifold point.  We are going to study the theory in the opposite, weakly coupled limit, but still at large $N$.

The extremal, supersymmetric black holes in the five dimensional asymptotically flat theory descend in this construction to the BTZ black holes of AdS$_3$ with periodic boundary conditions for fermions around the asymptotic circle in the AdS$_3$ geometry, i.e., they are in the Ramond sector of the theory.  The lightest black hole, which is massless, has the quantum numbers of a ground state in this sector. 

The construction of Ramond ground states of the D1-D5 CFT at the orbifold point is reviewed in detail in Appendix A of \cite{Balasubramanian:2005qu}; here we provide a brief summary.  We think of the CFT as describing  $N$ distinct ``strands'' of string, each of which propagates on $T^4$.   The ground states of the theory are formed by joining strands into various closed strings, which may be ``short'' (consisting of one strand) or ``long'' (consisting of multiple strands).  The strands are attached together by applying elementary bosonic ($\sigma$) and fermionic ($\tau$) twist operators which create $n$-wound string sectors.    Each twist operator has 8 polarizations associated with the global symmetries of the theory.   A general Ramond sector ground state is created by multiplying together bosonic and fermionic twist operators to achieve a total twist of $N = N_1 N_5$:
\begin{eqnarray}
\sigma &=& \prod_{n,\mu} (\sigma_n^\mu)^{N_{n\mu}} (\tau_n^\mu)^{N^{\prime}_{n\mu}}, \\
\sum_{n\mu} n (N_{n\mu} + N^{\prime}_{n\mu} )=  N,   & &~~ N_{n\mu} = 0,1,2,\ldots,~~~~ N^{\prime}_{n\mu} = 0,1,
\end{eqnarray}
where $\mu = 1,\ldots 8$ labels the polarizations, and $n = 1,2\ldots N$ labels the possible lengths of strings (i.e.\ the number of strands a string is made of). For our purposes, the integers $N_{n\mu}$ and $N^{\prime}_{n\mu}$, which count the various twist operators, 
uniquely specify a Ramond ground state of the theory.

  Note that while Ramond ground states all have the same energy, the spectrum of excitations around each of them is different.  For example, consider a case with $N=4$ strands.  Three possible states are four strings of length 1 ($N_1 = 4$), two long strings of length 2 ($N_2 = 2$), and one string of length 3 with another of length 1 ($N_3=1,N_1=1$).   If we take the CFT to be on a circle of circumference $L$, the momentum spectrum of excitations is very different in these sectors -- the first has modes gapped by $1/L$ with four-fold degeneracy, the second has modes gapped by $1/2L$ with 2-fold degeneracy and the third spectrum is a union of modes gapped by $1/L$ and $1/3L$ each with unit degeneracy.   Thus correlation functions computed in each of the microstates will be different and will depend on the twist distributions $\{N_{n\mu},N^{\prime}_{n\mu} \}$.

\begin{figure}[h!]
\centering
\includegraphics[width=0.5\textwidth]{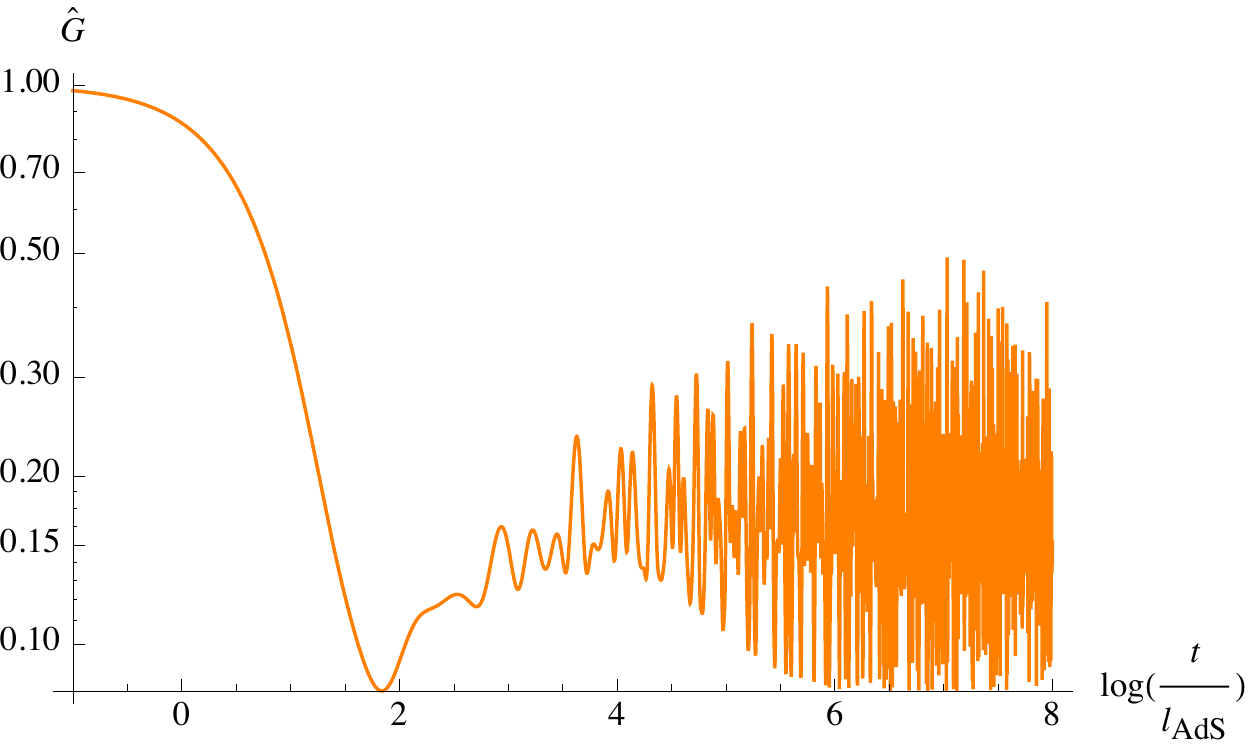}
\caption{The regularized two-point function (\ref{gencorr}). }
\label{fig:2pt}
\end{figure}

Now observe that the number of Ramond ground states is given by counting (colored) partitions of the integer $N$.   When $N\gg1$,  there are $O(e^{2\sqrt{2\pi N}})$  such partitions, leading to an enormous ground state degeneracy in the theory, corresponding to an entropy $S=2\sqrt{2\pi N}$.   Nearly all partitions of a large integer lie very close to a certain ``typical partition'' \cite{Balasubramanian:2005qu}.  This means that most Ramond states will in fact have twist distributions $\{N_{n\mu},N^{\prime}_{n\mu} \}$ that lie close to a certain typical distribution.   Thus, although correlation functions measured in individual microstates will depend on the precise form of the state, for almost all microstates the generic correlation functions will take a typical form, which we seek to investigate.  
Microcanonically, we should study all partitions of integers that lead to a total twist of $N$.    The easiest way, however, to derive the form of the typical state is to use the grand canonical ensemble with a ``chemical potential'' $\eta$ to fix the total ``charge'' $N$ for eight types ($\mu = 1 \cdots 8$) of bosons ($\sigma_n^\mu$) and fermions ($\tau_n^\mu$) with integral charges $n$.    When $N$ is large, the grand canonical average populations for $\{N_{n\mu},N^{\prime}_{n\mu} \}$ will also be typical, in the sense that most configurations will be very close to the average (the standard deviation over the mean will tend to zero).  Thus we can derive that most of the Ramond ground states have twist distributions close to the Bose-Einstein and Fermi-Dirac forms:
\begin{equation}
N_{n\mu} = {1 \over e^{\eta n} - 1}, ~~~~~~~ N^\prime_{n\mu} = {1 \over e^{\eta n} + 1}, ~~~~~~~ N_n = \sum_\mu (N_{n\mu} + N^\prime_{n\mu}) = {8 \over \sinh \eta n},
\label{typicaltwist}
\end{equation}
with $\eta$ set by
\begin{equation}
N = \sum_n n N_n \approx {2\pi^2 \over \eta^2}.
\label{temperature}
\end{equation}
For further reference, we note that the entropy scales as $S\sim 1/\eta$.

Now that we know the form of the typical Ramond ground state in the D1-D5 system it remains to calculate the correlation function.  Again following \cite{Balasubramanian:2005qu}, we will consider bosonic non-twist operators, which do not cut and join the $N$ strands of the CFT.  (An operator describing a fluctuation of the metric in the $T^4$ directions is an example.)  We focus on $S_N$ invariant operators obtained as sums of copies acting on each strand,
\begin{equation}
{\cal O} = {1 \over \sqrt{N}} \sum_{a=1}^N {\cal O}_a \, .
\label{operator}
\end{equation}
We are interested in two-point functions of the form
\begin{equation}
\langle \sigma^\dagger {\cal O}^\dagger {\cal O} \sigma \rangle \, .
\end{equation}
Since the state as a whole splits into a product of strings of lengths $n$, the correlator splits into a sum of terms, each of which reduces to a two point function in a CFT on a spatial circle that is $n$ times as long.  After some algebra (see \cite{Balasubramanian:2005qu} for details), the correlation function becomes
\begin{equation}
\langle \sigma^\dagger {\cal O}^\dagger {\cal O} \sigma \rangle  =
G(w,\bar w) = \frac{1}{N}\sum_{n=1}^N nN_n \sum_{k=0}^{n-1}  \frac{C}{\left[2n \sin \left( \frac{w-2\pi k}{2n}\right)\right]^{2h}  \left[2n \sin \left( \frac{\bar w-2\pi k}{2n}\right)\right]^{2\bar{h}}}  \, .
\end{equation}
Here $h$ and $\bar{h}$ are the left and right conformal dimensions,  $C$ is a constant, the sum on $n$ accounts for the contribution from strings of length $n$, and the sum on $k$ accounts for the placement of operators on different strands of a long string.     Also, $w = w_1 - w_2$ and $\bar{w} = \bar{w}_1 - \bar{w}_2$ are differences in the lightcone positions of the probe operators.  In Lorentzian signature we will set 
\begin{equation}
w = \phi - t,  ~~~~~~~~~~~~ \bar{w} = \phi + t,
\end{equation}
 where $\phi$ and $t$ are dimensionless angular and time coordinates in the CFT, normalized by setting the length of the spatial circle to be equal to $2\pi$.     

The correlator $G(w,\bar{w})$ exhibits physical lightcone divergences on the cylinder.  
We can regularize these divergences by dividing by the vacuum correlation function, which is fixed by conformal invariance.  Focusing, for definiteness, on operators with conformal dimensions $h=\bar{h}=1$ this results in \cite{Balasubramanian:2005qu}
\begin{equation}
\hat G(w,\bar w) = \frac{1}{N}\sum_{n=1}^N nN_n \sum_{k=0}^{n-1} \left( \frac{4 \sin \frac{w}{2}\sin \frac{\bar w}{2}}{2n \sin \left( \frac{w-2\pi k}{2n}\right)2n \sin \left( \frac{\bar w-2\pi k}{2n}\right)} \right)^2.
\label{gencorr}
\end{equation}
Setting $\phi = 0$, we can evaluate the  temporal correlation function numerically -- the result is plotted in Fig.~\ref{fig:2pt}.  We see a smooth initial decay followed by sporadic behavior, which is qualitatively similar to the behavior of observables in a single draw from a random matrix theory or in the SYK model before the average over disorder.

What is the origin of this sporadic behavior? As shown in Appendix~\ref{app}, the two-point function (\ref{gencorr}) for $\phi = 0$ receives contributions from frequencies of the form $m/n$, with $n$ an integer labeling the length of a component string (so $1\leq n\leq N$). This is a dense spectrum consisting of all rational numbers with denominators smaller than $N+1$.   The mixing of this large number of incommensurate frequencies produces the rapid late time oscillations.    A feature of the theory is that excitations on different long strings do not interact at the orbifold point.  This means that the smallest frequencies that occur in the two-point function are much larger than implied by the dense spectrum because no terms depend on the energy {\it differences} between excitations of strings of different lengths.  (Hence, while the spectrum contains excitations with energies $1/(N-1)$ and $1/N$ the two point function does not contain the difference $1/(N-1) - 1/N$.)   If we were dealing with a fully chaotic system, we would expect all the degeneracies in the spectrum to be broken, leading to exponentially small energy spacings.
   When we later analyze smooth, time-averaged versions of our two-point function, we will see that the relatively large frequency gap will cause the late-time plateau value to be reached earlier than would have been the case for smaller gaps.

One might wonder why the averaging over states we have performed by going to the grand canonical ensemble has not led to a smoothing of the sporadic behavior, in the way that ensemble averaging does for random matrix theory and SYK.  The reason is that every state in the ensemble has exactly the same spectrum, albeit with different degeneracies.  Our ensemble average therefore does not have the same effect as the averaging over different spectra that produces smoothing in random matrix theory and SYK.  Thus, to obtain smooth late-time behavior, we will need another way of coarse graining, to which we turn next.


\section{Random matrices and progressive time-averaging}\label{sec:random}
\label{sec:rmt}

The sporadic late-time fluctuations of the two-point correlation function (\ref{gencorr}) are reminiscent of similar behavior found in \cite{Cotler:2016fpe} for the SYK model.  There, smooth curves were obtained by averaging over random couplings. The main object of study in \cite{Cotler:2016fpe} was the spectral form factor
\beq
\label{eq:formfactor}
F_\beta(t)=\sum_{m,n}e^{-\beta(E_m+E_n)}e^{-i(E_m-E_n)t},
\eeq 
where the sum runs over all the eigenvalues $E_n$ of a Hamiltonian drawn from an ensemble. The spectral form factor displays sporadic late-time behavior, which can be smoothed by averaging over the emsemble of Hamiltonians. The main result of \cite{Cotler:2016fpe} is that the result agrees very well with the spectral form factor in random matrix theory, again after averaging over the random Hamiltonians. One motivation for studying the spectral form factor is  found in the spectral decomposition of the thermal two point function,
\beq
\langle O(t)O(0)\rangle \sim \sum_{m,n} |\langle m|O(0)|n\rangle|^2 \, e^{-\beta E_m+i(E_m-E_n)t}.
\eeq
It is believed that the late-time behavior is controlled by the phases $e^{i(E_m-E_n)t}$, so that some features of the two point function in this regime are captured by the spectral form factor \eqref{eq:formfactor}. Another motivation is that the spectral form factor is a more primitive quantity than two-point functions, in that it can be obtained directly from the partition function and does not require the introduction of operators. As a result, it can be straightforwardly studied in random matrix theory.

Our D1-D5 CFT has a definite Hamiltonian, so we cannot resort to disorder averaging for smoothing the sporadic late-time behavior of two-point functions. Is there any other meaningful way in which the late-time oscillations can be smoothed? A natural idea is to coarse-grain the correlator over time. One quickly notices, however, that averaging with any fixed time window either fails to remove the late-time oscillations or significantly distorts the early-time decay. This leads to the idea of using a time-window that grows with time, which we refer to as progressive time-averaging. In order to motivate a specific prescription we turn to random matrix theory.   We will ask  whether the well-known result of ensemble averaging could be alternatively obtained by progressive time-averaging applied to a single Hamiltonian drawn from the ensemble. We will find that this is the case for a time window that grows linearly with time, which will motivate our use of this procedure in the context of the D1-D5 CFT.

\begin{figure}[H]
\centering
\includegraphics[width=0.7\textwidth]{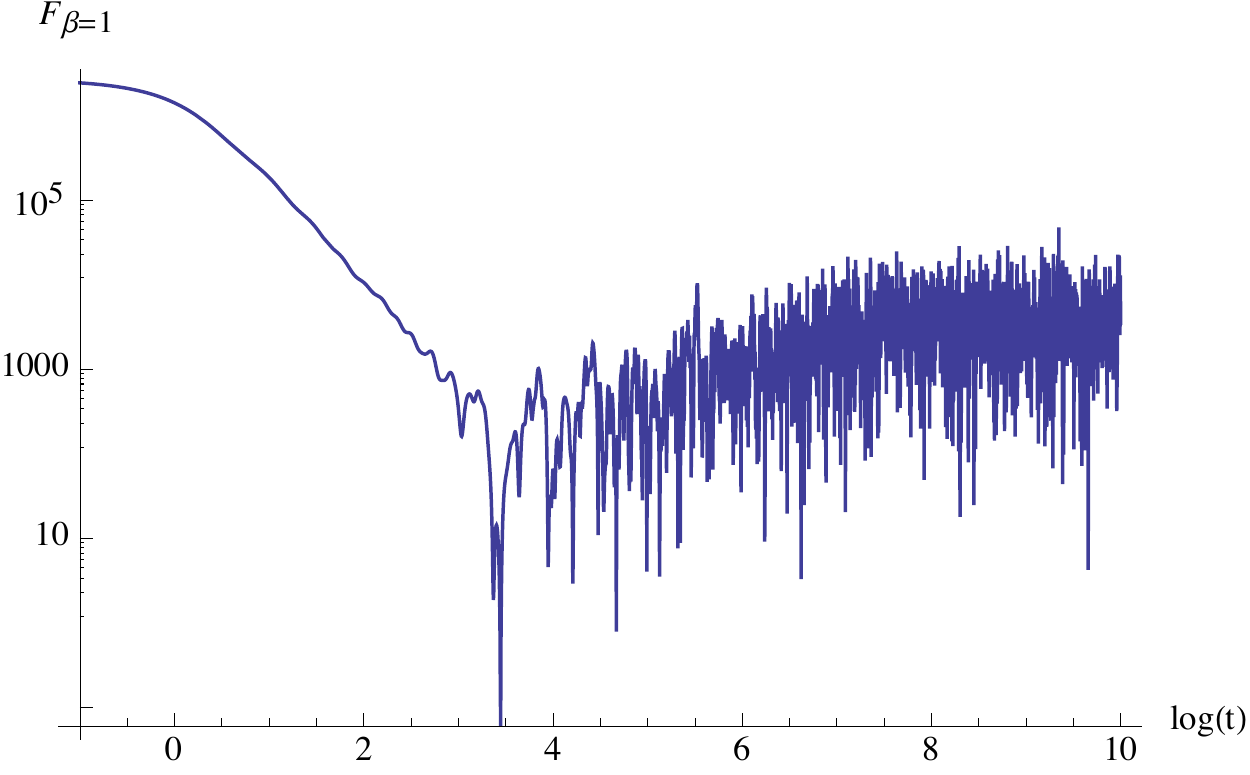}
\caption{Log-log plot of the spectral form factor \eqref{eq:formfactor} with $\beta=1$ for a single 1024$\times$1024 matrix drawn from the Gaussian Unitary Ensemble (GUE). The early part is self-averaging but the late part is superseded by noise.}
\label{fig:1}
\end{figure}

\subsection{Ergodicity in random matrix theory}

We consider Hamiltonians that are $L\times L$ matrices drawn from a random matrix ensemble. An important phenomenon in random matrix theory is self-averaging of certain quantities, i.e.\ the agreement of a quantity evaluated on a typical instance of the ensemble with the ensemble average of the same quantity. An interesting generalization of self-averaging quantities are ergodic quantities.
For ergodic quantities the result of averaging over random Hamiltonians can be approximately reproduced by using a single Hamiltonian drawn from the ensemble and coarse-graining in time.

It is known in random matrix theory that the spectral form factor is self-averaging for sufficiently short times but not for longer times \cite{prange1997spectral}. On the other hand, we can study the ergodicity of the form factor by considering suitable time averages
\beq
\bar F_{\beta}(t,\Delta t) = \int_{-\infty}^{\infty} dt' g(t-t',\Delta t) F_\beta (t'),
\eeq
where $g(t,\Delta t)$ is some smearing function with peak at $t=0$, width $\Delta t$ and $\int dt g(t,\Delta t)=1$. We could imagine it to be a Gaussian
\beq
g(t,\Delta t) = \frac{1}{\sqrt{2\pi} \Delta t}e^{-\frac{t^2}{2(\Delta t)^2}},
\eeq
or a step function, but its details should not matter too much.
The spectral form factor for a Gaussian random matrix ensemble, which is related to the late time behavior of the SYK model, is not ergodic for any fixed time window $\Delta t$~\cite{prange1997spectral}.\footnote{Note that for sufficiently large, fixed values of $\Delta t$ and $L$ the spectral form factor is ergodic for circular ensembles~\cite{haake1999fluctuations}.}

\begin{figure}[H]
\centering
\includegraphics[width=0.8\textwidth]{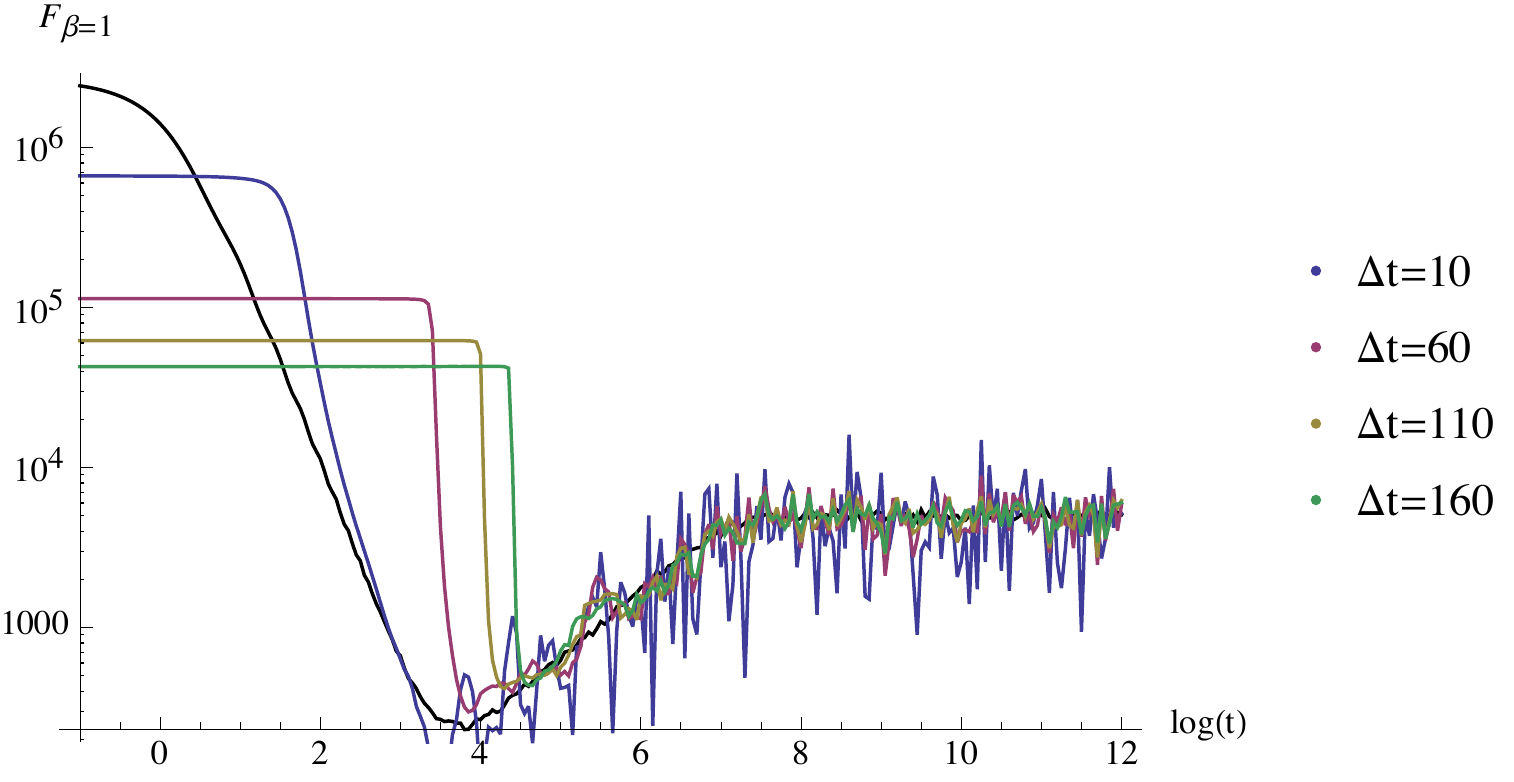}
\caption{Log-log plot of the average spectral form factor with $\beta=1$ for five hundred 1024$\times$1024 matrices drawn from the Gaussian unitary ensemble (GUE) (black), and the sliding window average \eqref{eq:slidingwindow} with fixed time windows $\Delta t =10,60,110,160$ (color) for a single instance of a random matrix. Notice that for averaging with a fixed time window there is tension between preserving the dip and having a sufficiently smooth ramp and plateau. }
\label{fig:2}
\end{figure}

\subsection{Progressive time-averaging}
We will now provide evidence suggesting that a progressive time average with $\Delta t \sim t$ gives a good approximation to the ensemble average for  Gaussian random matrices.  This is equivalent to averaging over fixed windows in $\log t$.
We first present a heuristic motivation, followed by numerical evidence.

In Gaussian random matrix ensembles, the probability distribution for the difference of two neighboring energy levels $s=E_{n+1}-E_n$ with $s>0$ is given by \cite{dyson1962statistical}
\beq
p(s)=\frac{A}{s_0} \left(\frac{s}{s_0}\right)^\beta e^{-\alpha \left( \frac{s}{s_0}\right)^2},
\eeq
where $s_0$ is the average value of $s$, $A$ is the normalization, and the constants $\alpha$ and $\beta$ depend on the specific ensemble.
In the spectral form factor (\ref{eq:formfactor}), the phases $\exp(\pm ist)$ appear with the same weight and so add to give a term proportional to $\cos(st)$.  Let us ask what happens to this cosine upon ensemble averaging: 
\beq
\label{eq:heurarg}
\int_0^\infty ds \, p(s) \cos(st) \sim \partial_t^\beta e^{-\frac{s_0^2 t^2}{4\alpha}}.
\eeq
We see a cancellation of the random phases in the average, resulting in Gaussian decay. Can we reproduce this with a time average? Consider for example a Gaussian smearing function applied to a typical phase $e^{is_0t}$ and its conjugate:
\beq
\int_{-\infty}^\infty dt' \frac{1}{\sigma} e^{-\frac{(t-t')^2}{2\sigma^2}}\cos(s_0 t') \sim \cos(s_0 t) e^{-\frac{s_0^2 \sigma^2}{2}}.
\eeq
We see that we need to set $\sigma = \frac{t}{\sqrt{2\alpha}}$ in order to reproduce the decay of the ensemble average. At the qualitative level the argument depends relatively little on the smearing function. For instance, for a step function we find
\beq
\frac{1}{\sigma}\int_{t-\sigma/2}^{t+\sigma/2} dt \cos(s_0t) = 2 \cos(s_0 t) \frac{\sin(\frac{s_0 \sigma}{2})}{s_0 \sigma},
\eeq
which for $\sigma \sim t$ is again a decaying function with width of order $\frac{1}{s_0}$.

\begin{figure}[H]
\centering
\includegraphics[width=0.9\textwidth]{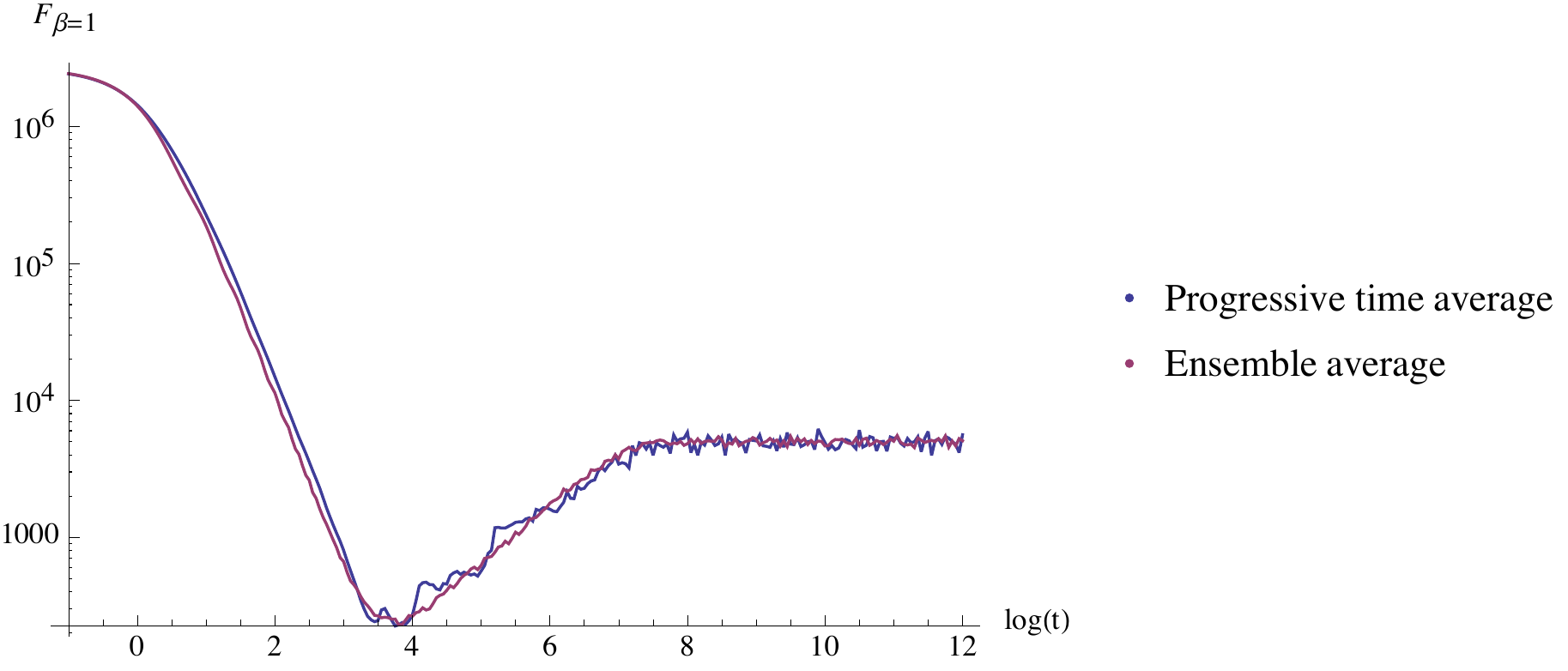}
\caption{Log-log plot of the average spectral form factor with $\beta=1$ for five hundred 1024$\times$1024 matrices drawn from the Gaussian unitary ensemble (GUE) (purple), and the sliding window average \eqref{eq:slidingwindow} for a progressive time window $\Delta t=0.8 t$ (blue) for a single instance of a random matrix. The progressive window captures the behavior of the ensemble average, in particular the dip, the ramp and the plateau.}
\label{fig:2prime}
\end{figure}

The argument above focuses on time dependences associated to differences of {\it neighboring} energy levels rather than generic energy differences, so that we may really trust it only at very late times. While it would be interesting to  find a more precise analytic argument, in the present paper we will simply use the heuristic argument as motivation, verify numerically that the resulting prescription produces good results in random matrix theory, and then apply it to our system of interest.

Now we present numerical evidence suggesting that the progressive time-average with window $\Delta t \sim t$ of the spectral form factor \eqref{eq:formfactor} is ergodic. We work with $1024\times 1024$ matrices and -- for numerical speed up -- discretized time averages
\beq
\label{eq:slidingwindow}
\bar F_{\beta}(t,\Delta t) = \frac{1}{100}\sum_{k=-50}^{49} F_{\beta}(t+\frac{k}{100}\Delta t).
\eeq 
We draw a single pseudorandom Hermitian matrix $H$ from the Gaussian Unitary Ensemble (GUE), $p(H)=\frac{1}{2^{L/2}\pi^{L^2/2}}e^{-\frac{L}{2}\text{Tr}H^2}$ with $L=2^{10}$. The spectral form factor for such a single matrix is plotted in Fig.~\ref{fig:1}. We see self-averaging for early times which is quickly overtaken by noise for late times.

The ensemble average of the spectral form factor for Gaussian ensembles is a well studied quantity. In particular, it behaves universally for large $L$, exhibiting an early slope from an initial value $\sim L^2$, followed by a dip, a linear rise over time scale $\sim L$, and finally a plateau which is the infinite-time average and is of order $L$ \cite{brezin1997spectral}. We plot this ensemble average for 500 random matrices in Fig.~\ref{fig:2} and Fig.~\ref{fig:2prime}, with the results of time-averaging for a single matrix drawn from GUE superposed, with various fixed windows and a progressive window $\Delta t=0.8t$, respectively.\footnote{The order one coefficient in front of $t$ should be smaller than 2  because otherwise, in (\ref{eq:slidingwindow}), we are calculating the total integral of the function.  Other than this constraint, the late part of the ramp and the plateau are rather insensitive to this coefficient, as follows from the argument presented around \eqref{eq:heurarg}. On the other hand, the location of the dip depends slightly on the choice of the coefficient.   The coefficient could be tuned to minimize deviations from the early self-averaging part of the curve in order extract the dip time.}
 It is clear that the progressive time window provides a much better approximation to the ensemble average than the fixed time windows.



\section{Echoes of chaos in D1-D5 two-point functions}
\label{sec:d1d5latetime}

Our examination of the D1-D5 theory at the orbifold point will focus on the regularized Lorentzian two-point function (\ref{gencorr}) evaluated at temporal distance $t$ and equal location in space:
\beq
\label{eq:reg2pt}
\hat G(t) = \frac{1}{N}\sum_{n=1}^N n N_n \sum_{k=0}^{n-1} \frac{\sin^4 \frac{t}{2}}{n^4 \sin^2 \left( \frac{t+2\pi k}{2n}\right)\sin^2 \left( \frac{t-2\pi k}{2n}\right)} \equiv
\frac{1}{N} \sum_{n=1}^N N_n C_n(t).
\eeq

\begin{figure}[H]
\centering
\includegraphics[width=.45\textwidth]{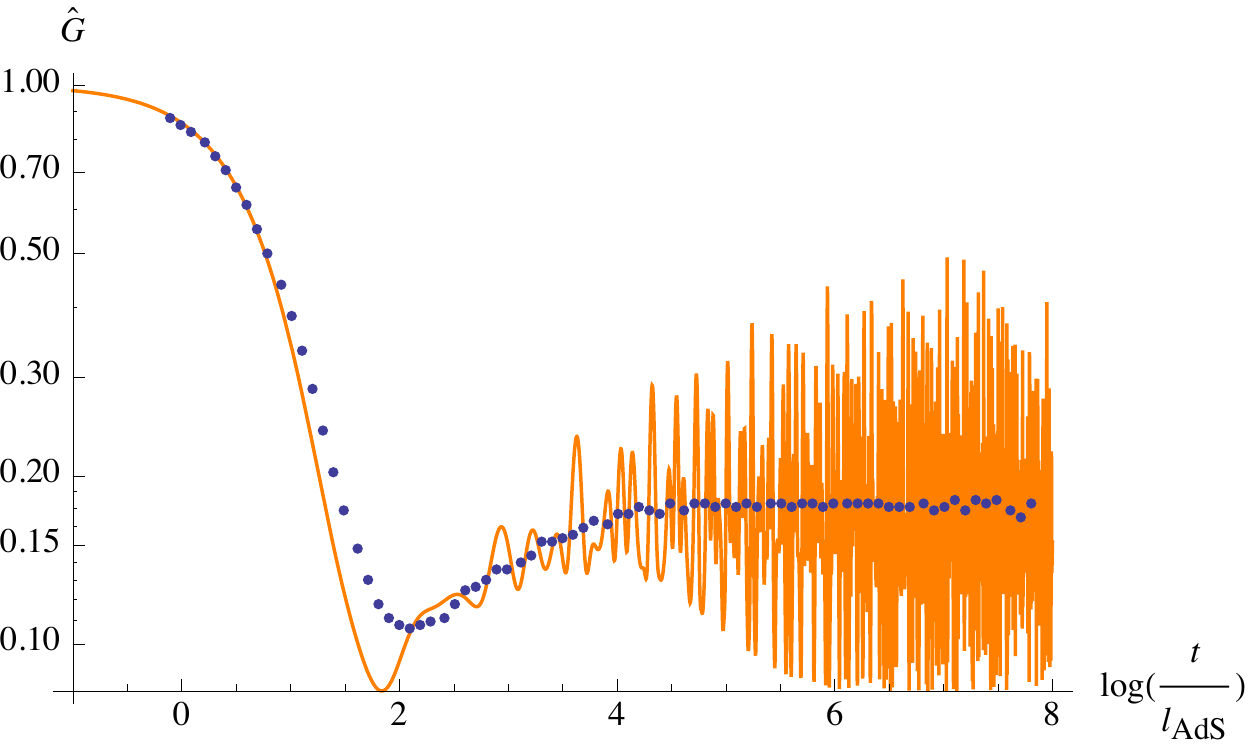}
\qquad
{\includegraphics[width=.45\textwidth]{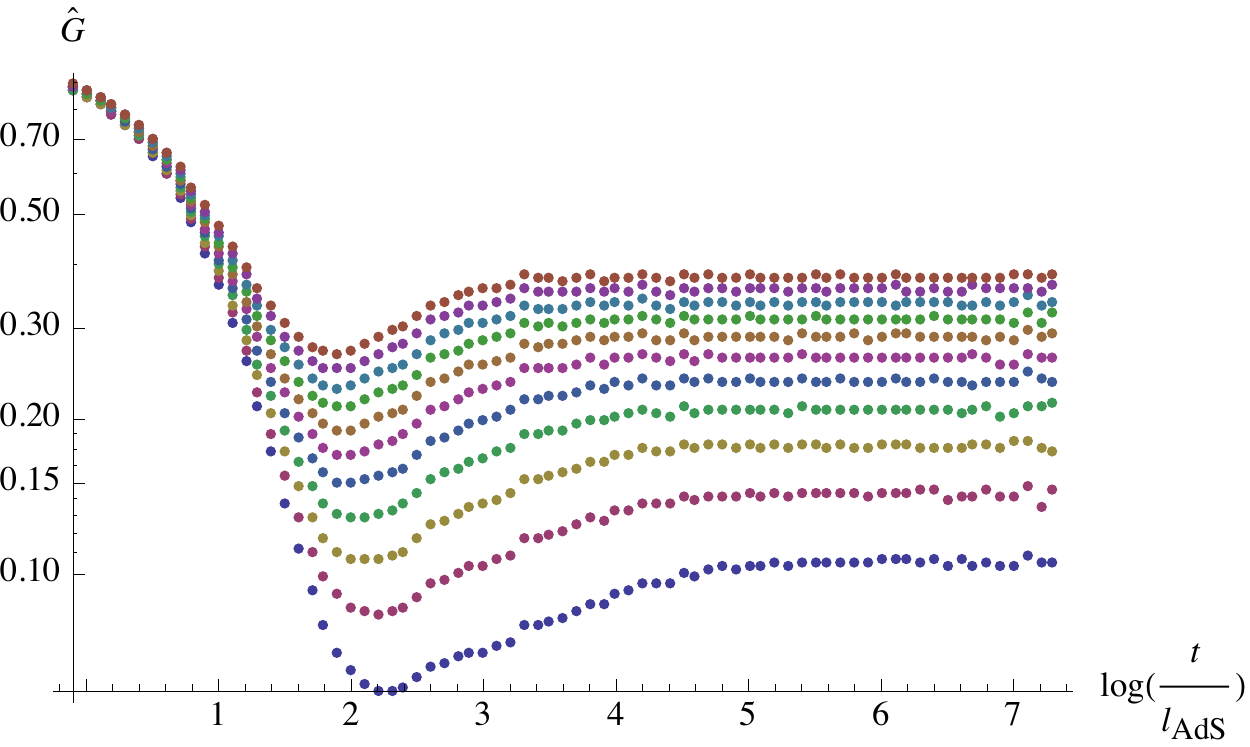}} \\
\caption{Left: The continuous orange line represents the regularized two-point function (\ref{eq:reg2pt}). The blue dotted line is its progressive time-average. Right: The progressive time-average of (\ref{eq:reg2pt}) for $\eta=0.05+0.025 j, \;\; j=0,...,10$. Smaller values of $\eta$ correspond to larger $N$ and smaller plateau height.}
\label{fig:d1d52pt}
\end{figure}

\noindent We start by applying the progressive time average of the previous section to this two point function. For numerical simplicity, we will use the pointwise averaging\footnote{One might expect a Gaussian kernel to produce smoother curves but it is numerically more challenging.} of \eqref{eq:slidingwindow} with $\Delta t=t$. The results are presented in Fig.~\ref{fig:d1d52pt}. The smoothed curve has a dip that is lower than the late time average (plateau), which it reaches after climbing a ramp whose length increases with $N$. This is in qualitative agreement with random matrix theory and the SYK model. The remainder of this section is devoted to an analytic study of the late time ramp and plateau, highlighting the quantitative differences from random matrix theory. 

\begin{figure}[H]
\centering
\includegraphics[width=0.5\textwidth]{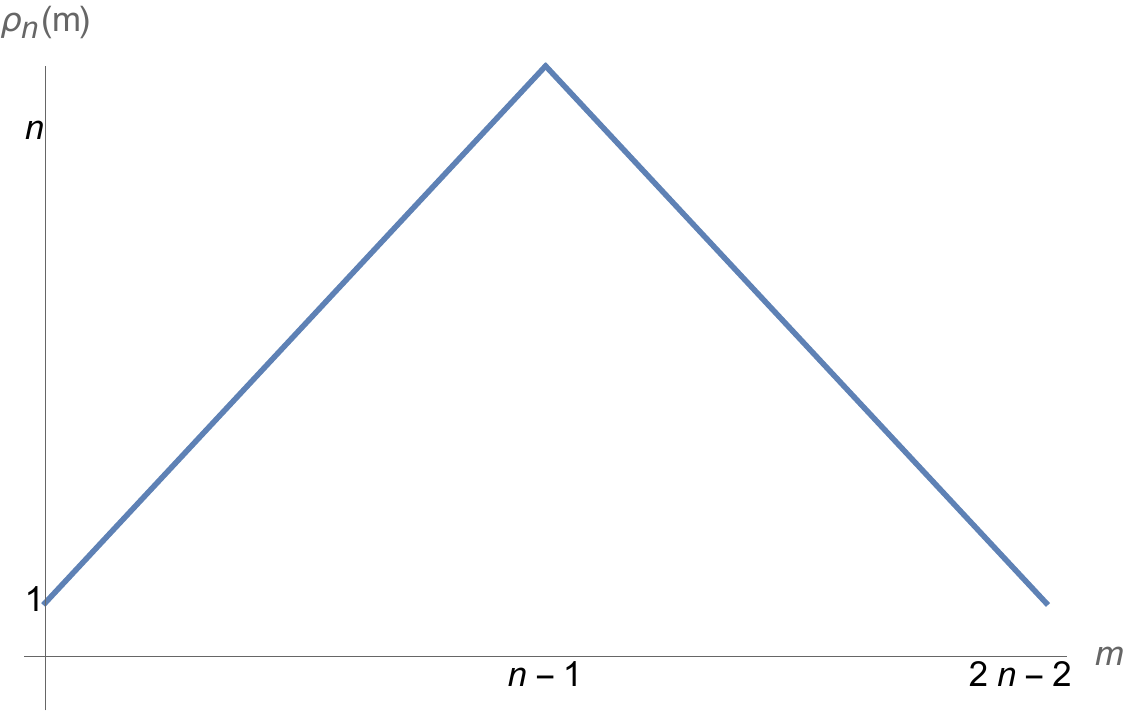}
\caption{The spectral weights \eqref{eq:spectdistr2} follow a `triangle law'.}
\label{fig:3}
\end{figure}

In the following, we will find two ways of rewriting $C_n(t)$ useful. The first was obtained in \cite{Balasubramanian:2005qu} by explicitly evaluating the sum over $k$:
\beq
C_n(t) = 2n\left(\frac{\sin \frac{t}{2}}{n\sin \frac{t}{n}}\right)^2 \left( 1+\frac{\sin t }{n\tan\frac{t}{n}}\right).
\label{simplebn}
\eeq
The second rewriting (worked out in  Appendix~\ref{app:2ptfunc}) will be handy for deriving analytically the late-time behavior of the correlator (\ref{eq:reg2pt}). In addition, it emphasizes the similarity between the D1-D5 two-point function and the spectral form factor: 
\beq\label{Bnbis}
C_n(t) = \frac{1}{n^3} \sum_{m_1=0}^{2n-2}\sum_{m_2=0}^{2n-2}\rho_n(m_1)\rho_n(m_2)\, e^{it \frac{(m_1-m_2)}{n}} \,\mathcal{G}_n(m_1+m_2+2).
\eeq
Here the `spectral weights' follow a `triangle law' (see Fig.~\ref{fig:3})
\beq
\label{eq:spectdistr2}
\rho_n(m) = \begin{cases} 
      m+1 & m< n \\
      2n-1-m & m \geq n \\
   \end{cases}
\eeq
and 
\beq
\label{gncases}
\mathcal{G}_n(x) = \begin{cases} 
      n & n~{\rm divides}~x \\
      0 & {\rm otherwise}. \\
   \end{cases}
\eeq
Below, we analyze the behavior of (\ref{eq:reg2pt}) after progressive time-averaging.

\subsection{Plateau}
At very late times, the progressive time averages of the quantities $C_n(t)$ tend to limiting values:
\begin{equation}
\bar C_n =\lim_{T\rightarrow \infty} \frac{1}{T}\int_{0}^T dt \,  C_n(t) = \frac{1}{n^3} \sum_{m=0}^{2n-2} \rho_n(m)^2\, \mathcal{G}_n(2m+2).
\end{equation}
The function $\mathcal{G}_n(2m+2)$ vanishes unless $2m+2$ is a multiple of $n$.  This requires that
\beq
m=\frac{n}{2}-1,\,n-1,\,\frac{3n}{2}-1
\eeq
for even $n$ and $m = n-1$ for odd $n$. This leads to
\begin{equation}
\bar C_{n({\rm even})} = \frac{3}{2} 
\qquad {\rm  and} \qquad
\bar C_{n({\rm odd})}= 1, 
\end{equation}
from which we obtain:
\begin{equation}
\bar{G}\equiv \lim_{T\rightarrow \infty} \frac{1}{T}\int_{0}^T dt \hat G(t) = {1 \over N} 
\left(
\frac{3}{2} \sum_{n \ {\rm even}}^N N_n  + \sum_{n \ {\rm odd}}^N N_n \right) = 
\frac{1}{N} \left(
\sum_{~{\rm all} \ n}^N N_n  + \frac{1}{2} \sum_{n \ {\rm even}}^N N_n \right).
\label{gfinalsum}
\end{equation}
We may approximate $\bar{G}$ for the typical state using the grand canonical ensemble in which the total twist $N$ is fixed with a chemical potential $\eta$ according to equations (\ref{typicaltwist}) and (\ref{temperature}).
In this approximation, we may let the sums in (\ref{gfinalsum}) run to infinity, 
\begin{equation}
\bar{G}\approx {1 \over N} \left( \sum_{s=1}^\infty {8 \over \sinh(\eta s)} + {1 \over 2} \sum_{s=1}^\infty {8 \over \sinh(2\eta s)} \right),
\end{equation}
and then, when $N$ is large, approximate them with integrals:
\begin{equation}
\bar{G} \approx 
\frac{1}{N} \cdot \frac{8}{\eta} \int_{\delta \eta}^\infty \frac{du}{\sinh u} 
+ 
\frac{1}{N} \cdot \frac{1}{2} \cdot \frac{8}{2 \eta} \int_{2 \delta \eta}^\infty \frac{du}{\sinh u} .
\end{equation}
Here $\delta$ is an $O(1)$ number that parameterizes the discretization error at the lower limit. Evaluating the integrals gives:
\begin{equation}
\bar{G} \approx {8\eta \over 2\pi^2} \left[ \log\coth\left({\eta \delta \over 2}\right) + {1\over 4} \log\coth\left(\eta \delta\right) \right].
\end{equation}
Remembering that $\delta= O(1)$ and $\eta=O(1/\sqrt{N})$, we further approximate and get:
\begin{equation}
\bar{G} \approx {5 \eta \over \pi^2} \log\left({1 \over \eta}\right) + O(1).
\label{plateausimpl}
\end{equation}
Thus, at late times, the coarse-grained temporal 2-point function approaches a constant plateau that scales as $\log(\sqrt{N})/\sqrt{N} \sim \log(S)/S$, where $S$ is the entropy of the black hole.

\subsection{Ramp}  
\label{sec:ramp}
To exhibit the ramp behavior of the correlator (\ref{eq:reg2pt}), we again employ progressive time averaging. For numerical convenience, we use a simple step-function averaging reminiscent of eq.~(\ref{eq:slidingwindow}):
\begin{equation}
\label{eq:2ptcoarsegrain}
\tilde{G}(t) = \frac{1}{t} \int_{t/2}^{3t/2} dt'\, \hat{G}(t').
\end{equation}
The progressively time-averaged $\tilde{G}(t)$ decomposes into progressively time-averaged contributions from individual modes,
\begin{equation}
\tilde{G}(t) = \frac{1}{N} \sum_n N_n \tilde{C}_n(t),
\end{equation}
each of which takes the form:
\begin{equation}
\label{eq:Ctilde}
\tilde{C}_n(t) = 
\frac{1}{t} \cdot \frac{2}{n} \int_{t/2}^{3t/2} 
\left( \frac{\sin t'/2}{\sin t'/n} \right)^2
\left( 1 + \frac{\sin t'}{n \tan t'/n}\right) dt'.
\end{equation}
This rewriting follows from eq.~(\ref{simplebn}). 

An intuitive approach to estimate the ramp is to recognize that for each individual $n$, the contribution of $\tilde{C}_n(t)$ to the correlator jumps fairly quickly from the low point in the dip to the plateau.  This is because there is only one timescale, as the gap (which sets the plateau time as $\gamma \times 1/{\rm gap}$ for some constant $\gamma$) and the level spacing are the same.\footnote{For further justification of this intuition and a slight improvement of the simple estimate below, see Appendix \ref{app:improvedramp}.}    We also know from the above that for odd $n$ the contribution to the plateau is $1$ and for even $n$ it is $3/2$.   This suggests that we can write the ramp part of the correlator as: 
\begin{equation}\label{rtsum}
\tilde{G}(t) \approx 
{1 \over N} 
\left(
\frac{3}{2} \sum_{n \ {\rm even}}^{t/\gamma} N_n  + \sum_{n \ {\rm odd}}^{t/\gamma} N_n \right)
=
\frac{1}{N} \left(
\sum_{~{\rm all} \ n}^{t/\gamma} N_n  + \frac{1}{2} \sum_{n \ {\rm even}}^{t/\gamma} N_n \right).
\end{equation}
In this equation we have taken into account that by a time $t$ the contribution of any $n$ with $n < t/\gamma$ will have hit its plateau, and we approximate the other modes as being $0$. 
Here $\gamma$ is some $O(1)$ number that relates the scale of the gap to the precise timescale of the plateau.  This may depend on the spectrum and on the operator.
The sum includes a unit contribution from the plateau for any $n$ and the second sum includes the additional $1/2$ that is present for even $n$.   Putting in the occupation numbers, we get
\begin{equation}
\label{eq:ramp}
\tilde{G}(t) = {1 \over N} \sum_{n=1}^{t/\gamma} {8 \over \sinh(\eta n)}   + {1 \over N} \sum_{n=1}^{t/(2\gamma)} {1 \over 2} {8 \over \sinh(2\eta n)}  .
\end{equation}
We plot this function on top of the time averaged two point function on Fig.~\ref{fig:4} with\footnote{To see better why this value works so well, see Appendix \ref{app:improvedramp}.} $\gamma=2$.
For large $N$ we again approximate the sum as an integral,
\begin{equation}
\sum_{n=1}^\tau {8 \over \sinh(\eta n)} \approx {8 \over \eta} \log\left[ {\tanh\left({\tau \eta \over 2}\right) \over \tanh\left({\delta\eta \over 2}\right)} \right].
\end{equation}
For the first sum $\tau = t / \gamma$ and for the second sum $\tau = t/(2\gamma)$.  Also, note that $\eta$ in second sum is multipled by a factor of 2.    Putting this all together and doing some elementary algebra gives
\begin{equation}
\tilde G(t) = {5 \eta \over \pi^2} \log\left[{1\over \eta \delta} \, \tanh\left({t \eta \over 2 \gamma }\right)\right] + {8 \eta \over 2 \pi^2} \log 2,
\label{finalrt}
\end{equation}
where we can ignore the last term  for our purposes. The characteristic time scale is the plateau time $t_p \sim 1/\eta \sim \sqrt{N}$.

Let us understand the time dependence.  First, note that $t$ cannot be taken to zero since the integral approximation is invalid in that case. Thus, $t$ is at least $\mathcal{O}(1)$ and $\tilde{G}(t)  \geq 0$.     We can consider two useful limits. First
\begin{equation}
\label{eq:largeNrampmidle}
1< t \ll \sqrt{N}: ~~~~~~~~ \tilde{G}(t) = {5 \eta \over \pi^2} 
\log\frac{t}{\gamma\delta},
\end{equation}
so the ramp rises logarithmically, in contrast to the linear rise for random matrices.   In this range,
\begin{equation}\label{earlyramp}
1< t \ll \sqrt{N}: ~~~~~~~~  {d\tilde{G} \over dt} \sim {\eta \over t} \sim {1 \over t  \sqrt{N}} .
\end{equation}
As a check also note that at late times
\begin{equation}
t \gg \sqrt{N}: ~~~~~~~~ \tilde{G}(t) = {5 \eta \over \pi^2}  \log{2 \over \eta \delta} \, ,
\label{limitplateau}
\end{equation}
which reproduces the plateau value from eq.~(\ref{plateausimpl}).

From (\ref{earlyramp}), we see that when $t \sim O(1)$, then $d\tilde{G}/dt \sim O(1/\sqrt{N})$, while when $t\sim O(\sqrt{N})$ at central times in the ramp, $d\tilde{G}/dt \sim 1/N$.  So the plateau height will be parametrically controlled by $d\tilde{G}/dt$ at central times in the ramp, multiplied by the duration of the ramp (which is determined by the inverse of the typical gap size, which is $1/\sqrt{N}$).  This gives the estimate $1/N \times \sqrt{N} \sim 1/\sqrt{N}$ for the plateau height.  The more rapid growth with slope of $O(1/\sqrt{N})$ in the early part of the ramp gives a logarithmic correction, resulting in a plateau height of $\log{N} /\sqrt{N} \sim \eta \log(1/\eta)$.  

\begin{figure}[H]
\centering
\includegraphics[width=0.6\textwidth]{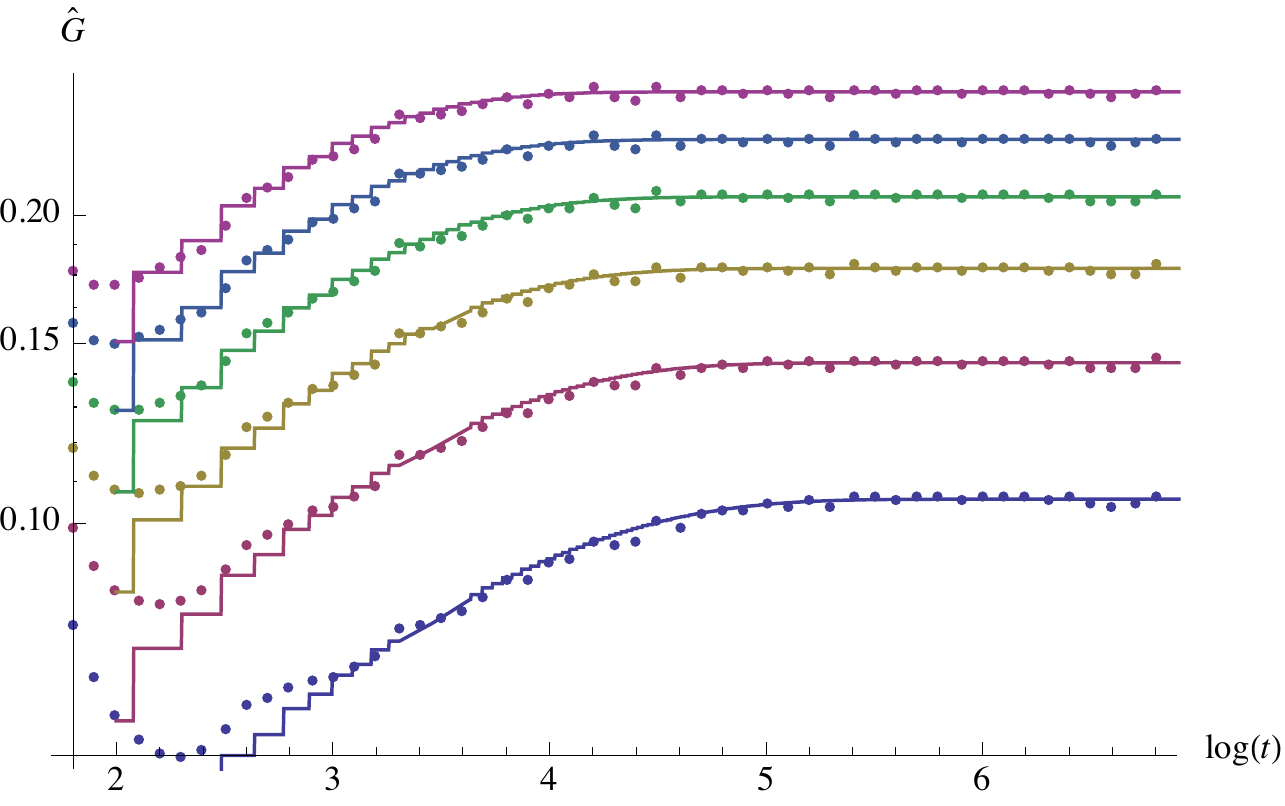}
\caption{The estimate \eqref{eq:ramp} for the ramp and the plateau with $\gamma=2$ (solid lines) versus the numerically evaluated progressive time averaged regularized two point function (dots) for $\eta=0.05,0.075,0.1,0.125,0.15,0.175$ (from bottom to top).}
\label{fig:4}
\end{figure}

\subsection{Dip}

In this subsection, we consider the temporal coarse graining of \eqref{eq:reg2pt} with generic progressive time window of width $\Delta t = a t$ (generalizing \eqref{eq:2ptcoarsegrain}), which we will denote by
\beq
\label{eq:generalcoarsegrain}
\tilde G_a(t) =  \frac{1}{a t} \int_{t-at/2}^{t+at/2} dt'\, \hat{G}(t')
\eeq
Such a generalization does not modify the conclusions about the late part of the ramp and the plateau time. However, we do expect the precise location of the dip to be sensitive to the parameter $a$ and therefore the specific coarse graining that we pick. As we will see, the scaling with the entropy is independent of $a$.

\begin{figure}[H]
\centering
\includegraphics[width=.45\textwidth]{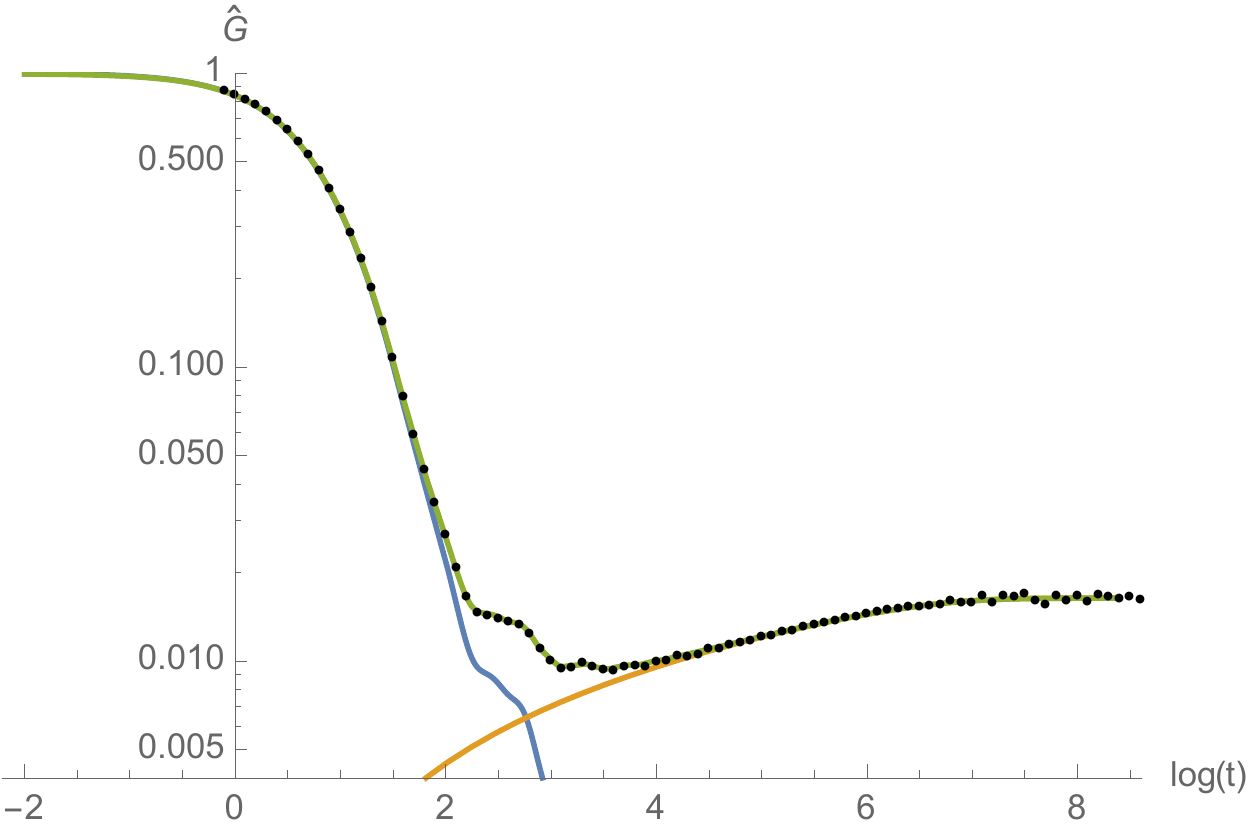}
\qquad
{\includegraphics[width=.45\textwidth]{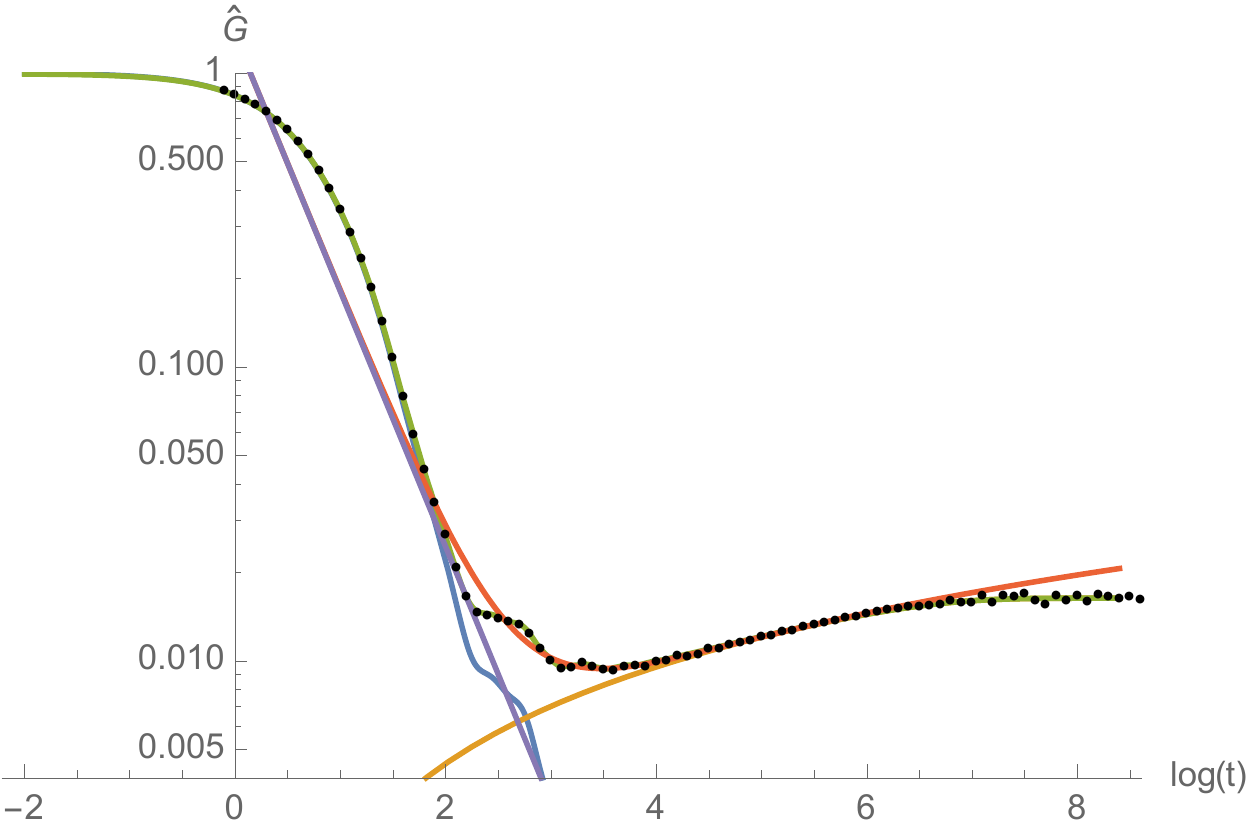}} \\
\caption{Left: The black dots represent the progressive time-average of the regularized two point function (\ref{eq:reg2pt}), the blue line is the progressive time-average of the BTZ two point function \eqref{eq:BTZreg}, the yellow line is the ramp estimate \eqref{eq:ramp}, while the green line is the sum of the latter two, i.e.\ the function \eqref{eq:sum}. Right: Same as the left, with the addition of: purple line is the BTZ asymptote \eqref{eq:BTZasym}; the dark orange line is the curve \eqref{eq:anotherestimate}. Both figures are for $\eta=0.005$ and use $\gamma=2$, $\delta=0.55$ and $a=1$.}
\label{fig:dip}
\end{figure}

The strategy we use is to approximate \eqref{eq:generalcoarsegrain} as the sum of the contribution coming from the regularized $M=0$ BTZ two point function 
\cite{Balasubramanian:2005qu}\footnote{To arrive at this expression, take (2.14) of \cite{Balasubramanian:2005qu}, set $w=-\bar w=-t$ and divide by the two point function in the NS vacuum.}
\beq
\label{eq:BTZreg}
\hat G_{BTZ}(t)=2\sin^2 \left( \frac{t}{2}\right)\frac{t+\sin t}{t^3}.
\eeq
and our ramp estimate \eqref{finalrt}, i.e.
\beq
\label{eq:sum}
\tilde G_a(t) \approx \frac{1}{a t} \int_{t-at/2}^{t+at/2} dt'\, \hat G_{BTZ}(t') +   {5 \eta \over \pi^2} \log\left[{1\over \eta \delta} \, \tanh\left({t \eta \over 2 \gamma }\right)\right]+ {8 \eta \over 2 \pi^2} \log 2 .
\eeq
We illustrate on the left of Fig. \ref{fig:dip} how remarkably well this naive estimate works. The very precise match indicates that there is no extra physics going on at intermediate time scales.

In order to extract the dip time, we want to find the minimum of this curve. Let us assume that the dip happens at $1\ll t_d \ll \sqrt{N}$. In this case, we can approximate the ramp part with the logarithmic rise of \eqref{eq:largeNrampmidle}. On the other hand, the BTZ part asymptotes to\footnote{For $a\geq 2$ the asymptote crosses over to $1/t$. In this case, the lower end of the integral moves to the left, so we no longer have a good temporal coarse graining.}
\beq
\label{eq:BTZasym}
\frac{1}{a t} \int_{t-at/2}^{t+at/2} dt'\, \hat G_{BTZ}(t') \approx \frac{4}{4-a^2} \frac{1}{t^2}, \qquad t\gg 1, \; a<2, 
\eeq
so that for times $1\ll t \ll \sqrt{N}$ we have
\beq
\label{eq:anotherestimate}
\tilde G_a(t) \approx \frac{4}{4-a^2} \frac{1}{t^2} + {5 \eta \over \pi^2} 
\log\frac{t}{\gamma\delta},
\eeq
see the right panel of Fig. \ref{fig:dip} for an illustration.
The minimum of this curve is at
\beq
t_d = \sqrt{\frac{8\pi^2}{5(4-a^2) \eta}},
\eeq
which scales as $t_d \sim \sqrt{S}$ with the entropy. Note that we indeed have $1\ll t_d \ll \sqrt{N}$. This establishes a parametrically long ramp. Notice that we have $t_d \sim \sqrt{t_p}$ which is also valid for random matrices.

\subsection{Variances}
Eqs.~(\ref{plateausimpl}) and (\ref{eq:ramp}) apply to the typical Ramond ground state of the D1-D5 system. We may ask by how much other Ramond ground states differ from these typical values. 

We start by considering eigenstates of the occupation numbers $N_n$. The middle expression in eq.~(\ref{rtsum}) casts $\tilde{G}(t)$ as a linear combination of distinct occupation numbers $N_n$, which in the grand canonical ensemble are independent random variables. Thus, the variance in the ramp part of progressively time-averaged correlator, $\tilde{G}(t)$, can be approximated in terms of variances in $N_n$:
\begin{equation}
\textrm{var}\,\tilde{G}(t) \approx \frac{1}{N^2} \left(
\frac{9}{4} \sum_{n~\textrm{even}}^{t/\gamma} \textrm{var}\,N_n 
+ 
\sum_{n~\textrm{odd}}^{t/\gamma} \textrm{var}\,N_n \right).
\end{equation}
Using
\begin{equation}
\textrm{var}\,N_{n\mu} = \frac{e^{\eta n}}{(e^{\eta n} - 1)^2} \qquad {\rm and} \qquad
\textrm{var}\,N'_{n\mu} = \frac{e^{\eta n}}{(e^{\eta n} + 1)^2}\,,
\end{equation}
we get:
\begin{equation}
\textrm{var}\,N_n = \frac{8 \cosh \eta n}{\sinh^2 \eta n}.
\end{equation}
Substituting, we obtain:
\begin{align}
\textrm{var}\,\tilde{G}(t) & \approx \frac{8}{N^2} \left( 
\frac{1}{\eta} \int_{\eta\delta}^{\eta t / \gamma} 
\frac{du \cosh u}{\sinh^2 u}
+
\frac{5}{4} \cdot \frac{1}{2\eta} \int_{2\eta\delta}^{\eta t / \gamma} 
\frac{du \cosh u}{\sinh^2 u}
\right) \nonumber \\
& = \frac{8}{N^2\eta} \left(
\frac{1}{\sinh \eta \delta} + \frac{5}{8 \sinh 2 \eta \delta}
- \frac{13}{8\sinh \eta t / \gamma}
\right).
\end{align}
The variance in the plateau height is obtained by taking the late time limit of the above expression:
\begin{equation}
\textrm{var}\,\tilde{G}(\infty) \approx
\left(\frac{\eta^2}{2\pi^2}\right)^2\cdot\frac{8}{\eta}
\left( \frac{1}{\sinh \eta \delta} + \frac{5}{8 \sinh 2 \eta \delta}
\right)
\propto \eta^2,
\end{equation}
where we have expanded in small $\eta$ (large $N$) in the last expression.
So the standard deviation in the plateau height divided by the mean (\ref{plateausimpl}) scales as $1/\log(1/\eta) \sim 1 / \log N$.
 Note that the primary sources of the deviation are the `relatively short long strings' with $n \gtrsim 1$.

We can also estimate the variance of the slope of the ramp.
From (\ref{rtsum}), we recognize that after sufficient coarse-graining, $d\tilde{G}(t)/dt \propto N_{t/\gamma}/N$. Therefore, 
\begin{equation}
\textrm{var}\left(\frac{d\tilde{G}(t)}{dt}\right) \propto 
\frac{{\rm var}\,N_{t/\gamma}}{N^2} \propto
\frac{\eta^4 \cosh (\eta t / \gamma)}{\sinh^2 (\eta t / \gamma)}.
\label{coarse-variance}
\end{equation}
At central times in the ramp $t \sim \sqrt{N}$.  Since $\eta \sim 1/\sqrt{N}$ this means that the hyperbolic functions on the right hand size are $O(1)$.  So the standard deviation in the slope is $O(1/N)$.  At central times in the ramp, we showed above that the slope is also $O(1/N)$.  Thus the slope varies significantly between occupation number eigenstates.

However, a typical Ramond ground state is a {\it superposition} of these eigenstates. Since $\tilde{G}(t)$ is not an eigenvalue, but an expectation value of a quantum mechanical operator, it makes sense to discuss the variances among all ensemble members, including superpositions. Such a variance, with superposition states weighted with a uniform measure over $\mathbb{CP}^{\,\exp S}$, was computed in \cite{Balasubramanian:2007qv}.   This paper showed that the variance in the expectation value of a quantum mechanical operator is suppressed relative to the variance among its eigenstates by an extra factor of the dimension of the Hilbert space. Thus, all the variances computed in the preceding paragraphs receive an additional factor:
\begin{equation}
{\rm var}_{\rm superpositions} = 
e^{-S}\, {\rm var}_{N_n{\rm eigenstates}} =
e^{-2\pi\sqrt{2N}}\, {\rm var}_{N_n{\rm eigenstates}}\,.
\end{equation}
Thus we can conclude that over the entire Hilbert space, almost all states will show a coarse-grained two point function that lies very close to the results that we have computed in the typical state.


\section{Discussion}
\label{sec:disc}

We have studied the time-ordered two-point correlation function of certain operators in typical states of the Ramond sector of the D1-D5 CFT.   At strong coupling these are black hole microstates and the theory is expected to be chaotic.  Here, we studied the weak coupling limit of this theory, where it is integrable.  After temporal coarse-graining, the late time two-point function displays a characteristic dip, ramp and plateau.  These features are remarkably similar to those seen in random matrix theory (RMT) and the SYK model, showing that the qualitative form does not specifically arise from the chaos present in those models.

A key quantitative difference is that the slopes of the ramps in RMT and SYK are constant, while in our model the slope decreases with time.  Also, while the RMT and SYK plateaus are exponentially suppressed in the entropy $S$,  our plateau scales as $\log S / S$.  Finally, the plateau in RMT and SYK is reached at a time that is exponential in the entropy, while in our case it is reached at times proportional to the entropy.   

 These quantitative differences arise from the different structures of the excitation spectra.  In a chaotic theory the energy eigenvalues are typically non-degenerate and have spacings that are exponentially small in the entropy.   Random matrix theories also demonstrate a phenomenon of spectral rigidity, in which repulsion between eigenvalues of the Hamiltonian produces long-range correlations in the spectrum.  The exponentially small gap leads to the exponentially large plateau time, and the linear ramp is partially a consequence of the spectral rigidity \cite{Cotler:2016fpe}.   By contrast, although the D1-D5 theory at the orbifold point has a dense spectrum, there is a very large degeneracy of each energy level and the gaps are not exponentially small.    This leads to a much shorter timescale for the plateau.  The plateau is also much higher because the theory explores its phase space less completely than a chaotic model.

The authors of \cite{Dyer:2016pou} argue that in general 2d CFTs, the dip occurs at times proportional to the entropy.  Likewise the authors of \cite{Fitzpatrick:2016ive} predict a breakdown of the semiclassical description of the two point function at entropy times. (See also \cite{Galliani:2016cai} for related work in the context of the D1-D5 system.)  In contrast, the location of our dip scales with the entropy as $\sqrt{S}$. The reason for the difference is that all of these works address finite temperature states, and require generalization to apply to the zero temperature, large entropy system that we examine.\footnote{In SYK, the range of parameters where there is both chaotic behavior and IR conformal symmetry is $1\ll\beta J\ll N$, where $\beta$ is the inverse temperature and $J$ is the coupling \cite{Maldacena:2016hyu}. We see that in the zero temperature limit we need to switch off the coupling to stay in this regime. Our situation is somewhat similar to this.}   It would also be useful to see what the results in \cite{Dyer:2016pou,Fitzpatrick:2016ive} imply for the late time, finite temperature two-point function in the orbifold D1-D5 theory.

It would be very interesting to see how these phenomena change as the D1-D5 theory is deformed from the integrable point that we studied to the strongly coupled region where it is expected to be chaotic and dual to weakly coupled AdS$_3$ gravity.    One strategy for making progress is to turn on this marginal deformation perturbatively \cite{Carson:2016uwf}, although it may be challenging to sum the perturbation series with sufficient accuracy to capture the late time physics.    Another interesting avenue is to consider correlation functions of twist operators that induce interaction between the long string components of the state.  The resulting mixing should break degeneracies between energy levels and lead to much smaller gaps.   This will in turn lead to much longer timescales for the ramp and the plateau in the two point function.

\section*{Acknowledgments}

We thank Pawe\l{} Caputa, Federico Galli, Aitor Lewkowycz, Alex Maloney, M\'ark Mezei, Onkar Parrikar, Charles Rabideau and Douglas Stanford for helpful discussions.   This work was supported in part by a grant from the Simons Foundation (\#385592, Vijay Balasubramanian) through the It From Qubit Simons Collaboration, by the U.S. Department of Energy under contract DOE DE-FG02-05ER- 41367, by the Belgian Federal Science Policy Office through the Interuniversity Attraction Pole P7/37, by FWO-Vlaanderen through projects G020714N and G044016N, and by Vrije Universiteit Brussel through the Strategic Research Program ``High-Energy Physics''. The work of B. Czech is supported by the Peter Svennilson Membership in the Institute for Advanced Study.


\appendix

\section{D1-D5 two-point functions echo the spectral form factor}\label{app}
\label{app:2ptfunc}

To highlight the similarity between
\beq
\label{gtappendix}
\hat G(t) = \frac{1}{N}\sum_{n=1}^N n N_n \sum_{k=0}^{n-1} \frac{\sin^4 \frac{t}{2}}{n^4 \sin^2 \left( \frac{t+2\pi k}{2n}\right)\sin^2 \left( \frac{t-2\pi k}{2n}\right)} \equiv
\frac{1}{N} \sum_{n=1}^N N_n C_n(t)
\eeq
and the spectral form factor, we start by noting that
\beq
\frac{\sin \frac{t}{2}}{\sin \left( \frac{t-2\pi k}{2n}\right)} = (-1)^k\frac{q^n-q^{-n}}{q-q^{-1}}= (-1)^k\sum_{\ell=0}^{n-1} q^{2\ell+1-n} = (-1)^k\sum_{\ell=0}^{n-1}e^{i\frac{t-2\pi k}{2n} (2\ell+1)-i \frac{t-2\pi k}{2}},
\eeq
where $q = \exp\left(i \frac{t - 2 \pi k}{2n} \right)$. Substituting into \eqref{gtappendix} yields
\beq
C_n(t) = \frac{1}{n^3} \sum_{\ell_1,\ell_2,\ell_1',\ell_2'=0}^{n-1}e^{i\frac{\ell_1+\ell_1'-\ell_2-\ell_2'}{n}t} \,
\mathcal{G}_n(\ell_1+\ell_1'+\ell_2+\ell_2'+2)
\label{cninterm}
\eeq
with 
\bea
\mathcal{G}_n(x) &=\sum_{k=0}^{n-1}e^{2\pi i(2-\frac{x}{n})k} = \frac{e^{-2\pi i x}-1}{e^{-2\pi i \frac{x}{n}}-1}\,.
\eea
Because $x$ only takes integer values in (\ref{cninterm}), we effectively have
$\mathcal{G}_n(x) =\sum_{q\in \mathbb{Z}}n\delta_{qn,x}$, which is eq.~(\ref{gncases}) in the main text.

Since the summand in (\ref{cninterm}) depends only on the combinations $m_1=\ell_1+\ell_1'$ and $m_2=\ell_2+\ell_2'$, we change two of the summation variables to these new variables. 
We exchange the sums according to
\beq
\sum_{\ell_1=0}^{n-1} \sum_{m_1=\ell_1}^{\ell_1+n-1} =\sum_{m_1=0}^{n-1}\sum_{\ell_1=0}^{m_1} + \sum_{m_1=n}^{2n-2} \sum_{\ell_1=m_1-n+1}^{n-1} 
= \sum_{m_1=0}^{n-1}(m_1+1) + \sum_{m_1=n}^{2n-2} (2n-1-m_1),
\eeq
where in the last step we perform the $\ell_1$ sums as the (suppressed) summand does not depend on it. After a similar manipulation for $\ell_2$ and $m_2$, we find
\beq
C_n(t) = \frac{1}{n^3} \sum_{m_1=0}^{2n-2}\sum_{m_2=0}^{2n-2}\rho_n(m_1)\rho_n(m_2)\, e^{it \frac{(m_1-m_2)}{n}}\, \mathcal{G}_n(m_1+m_2+2),
\eeq
where the spectral weights are given by
\beq
\label{eq:spectdistr}
\rho_n(m) = \begin{cases} 
      m+1 & m< n \\
      2n-1-m & m \geq n. \\
   \end{cases}
\eeq
We plot this function in Fig.~\ref{fig:3}. 

\section{More detailed ramp estimate}
\label{app:improvedramp}

\begin{figure}[h]
\centering
\includegraphics[width=.35\textwidth]{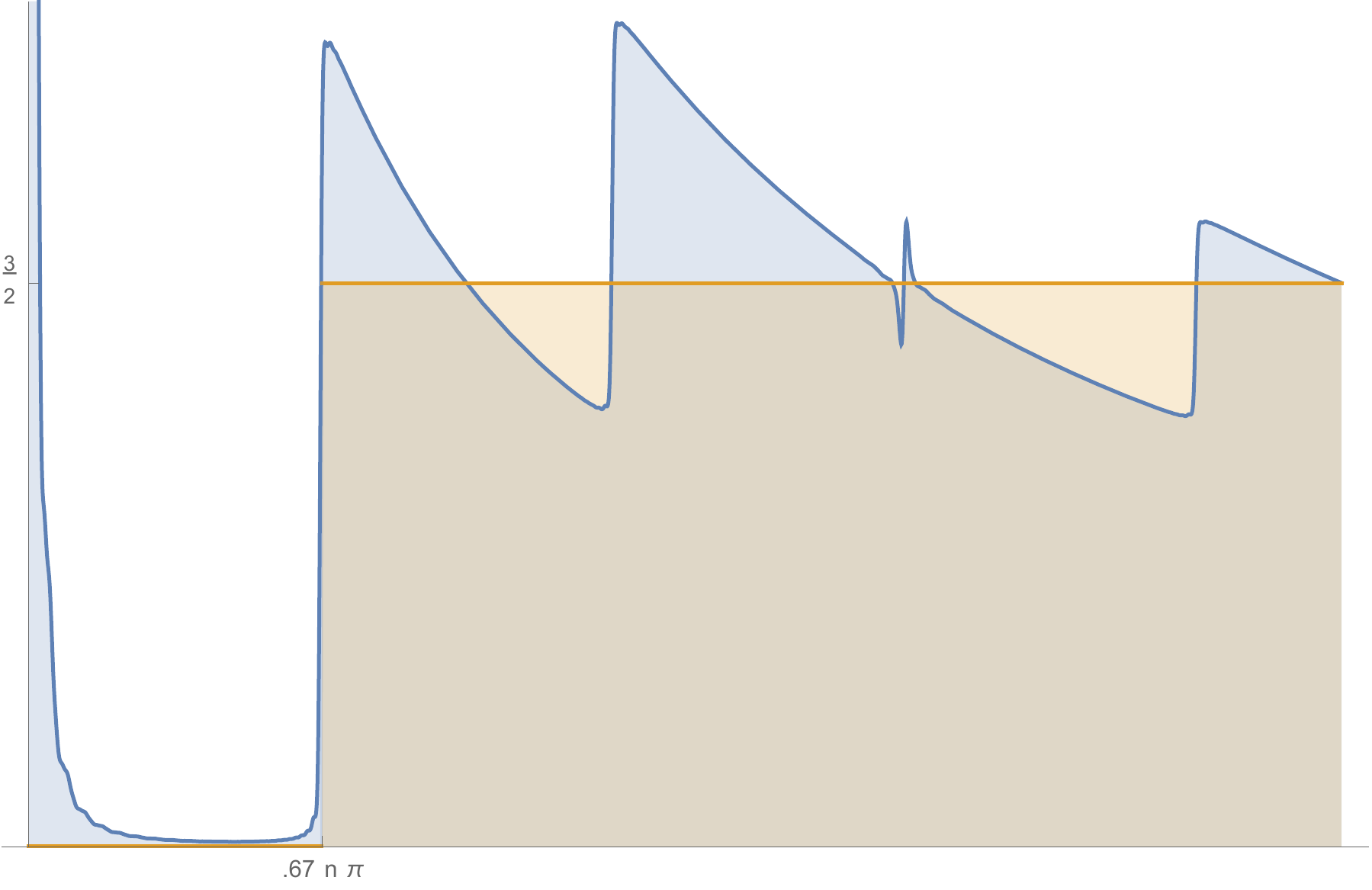}
\qquad\qquad
{\includegraphics[width=.35\textwidth]{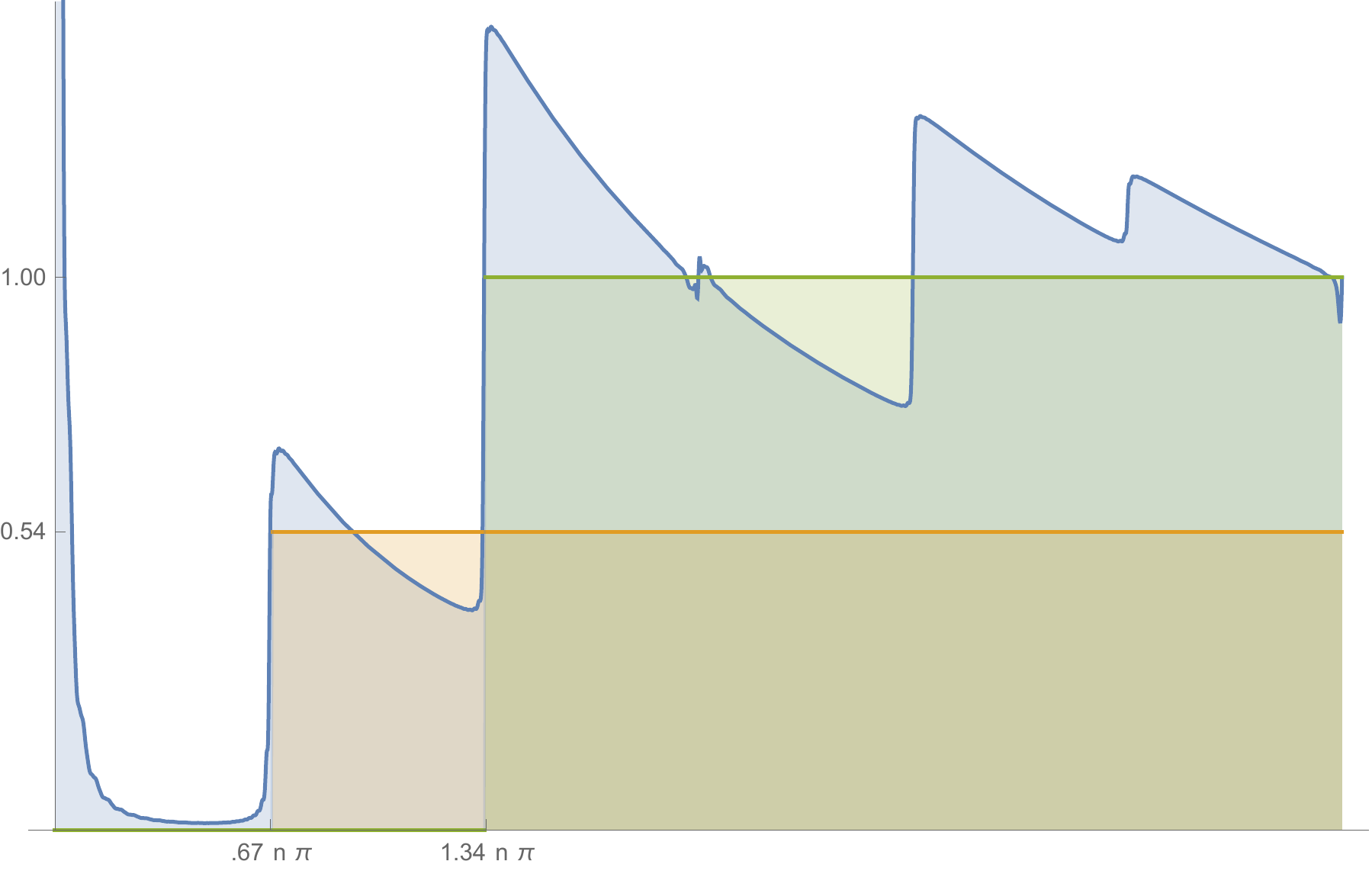}} \\
\vspace*{5mm}
{\includegraphics[width=.35\textwidth]{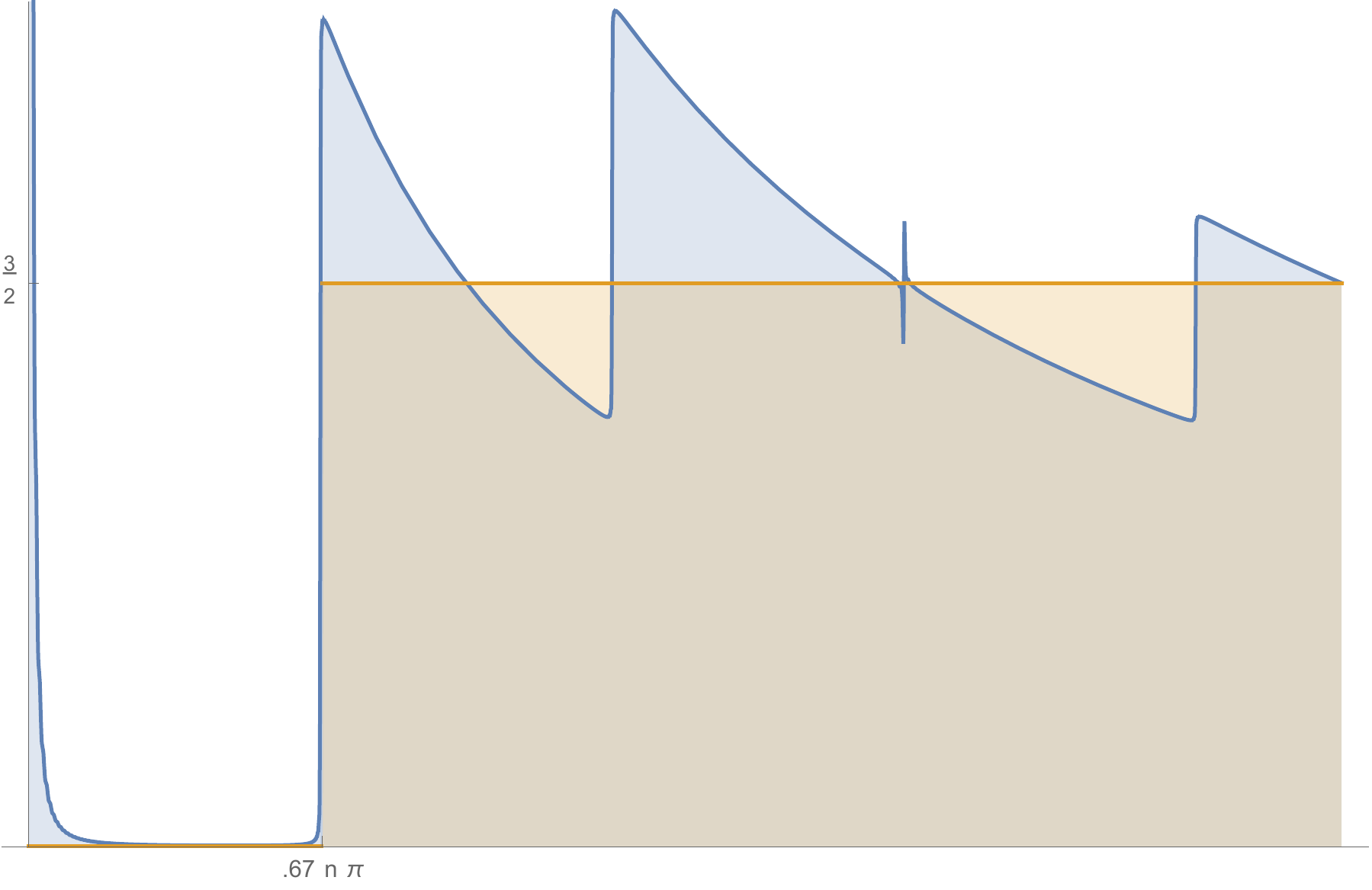}}
\qquad\qquad
{\includegraphics[width=.35\textwidth]{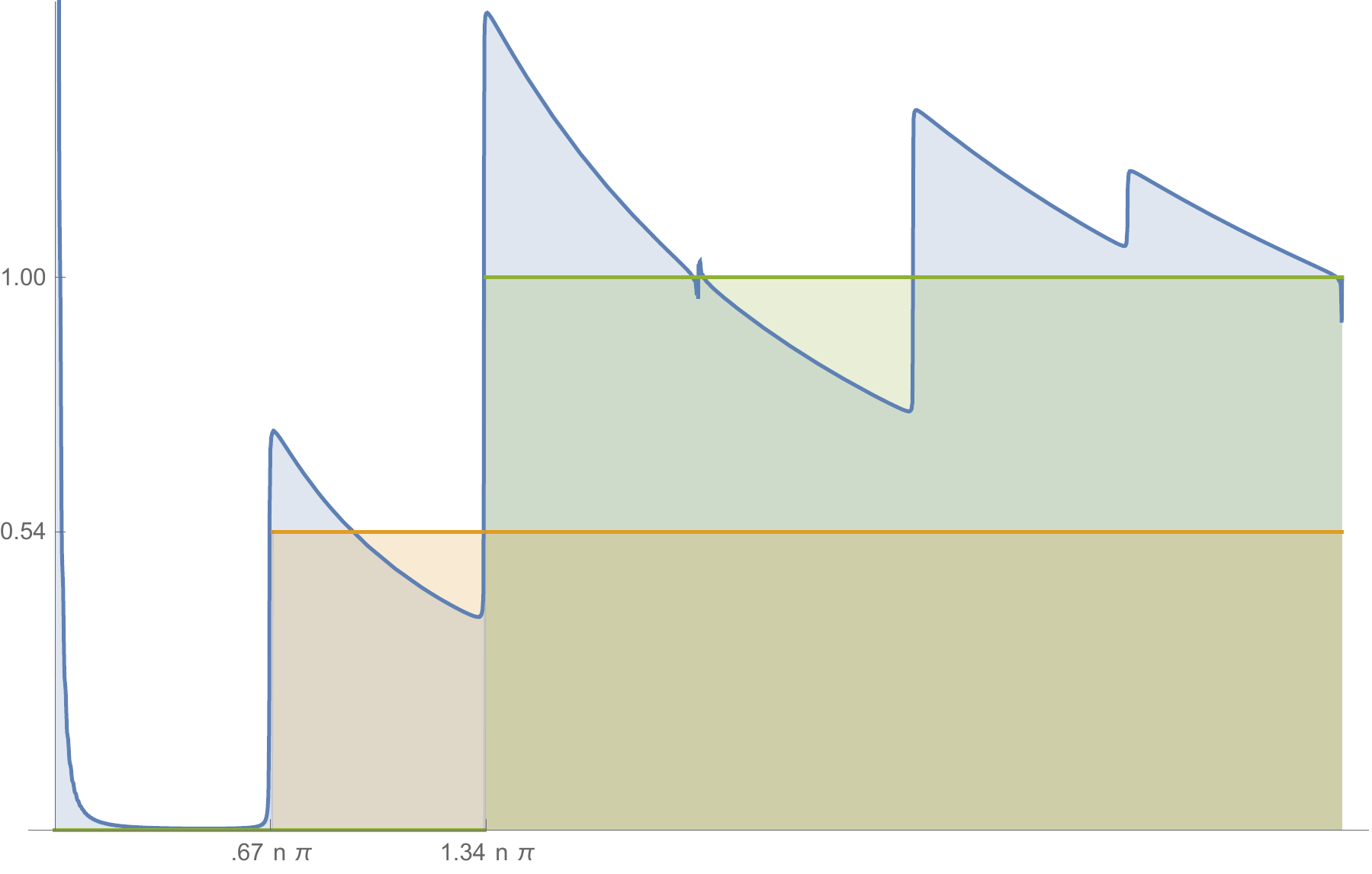}}\\
\vspace*{5mm}
{\includegraphics[width=.35\textwidth]{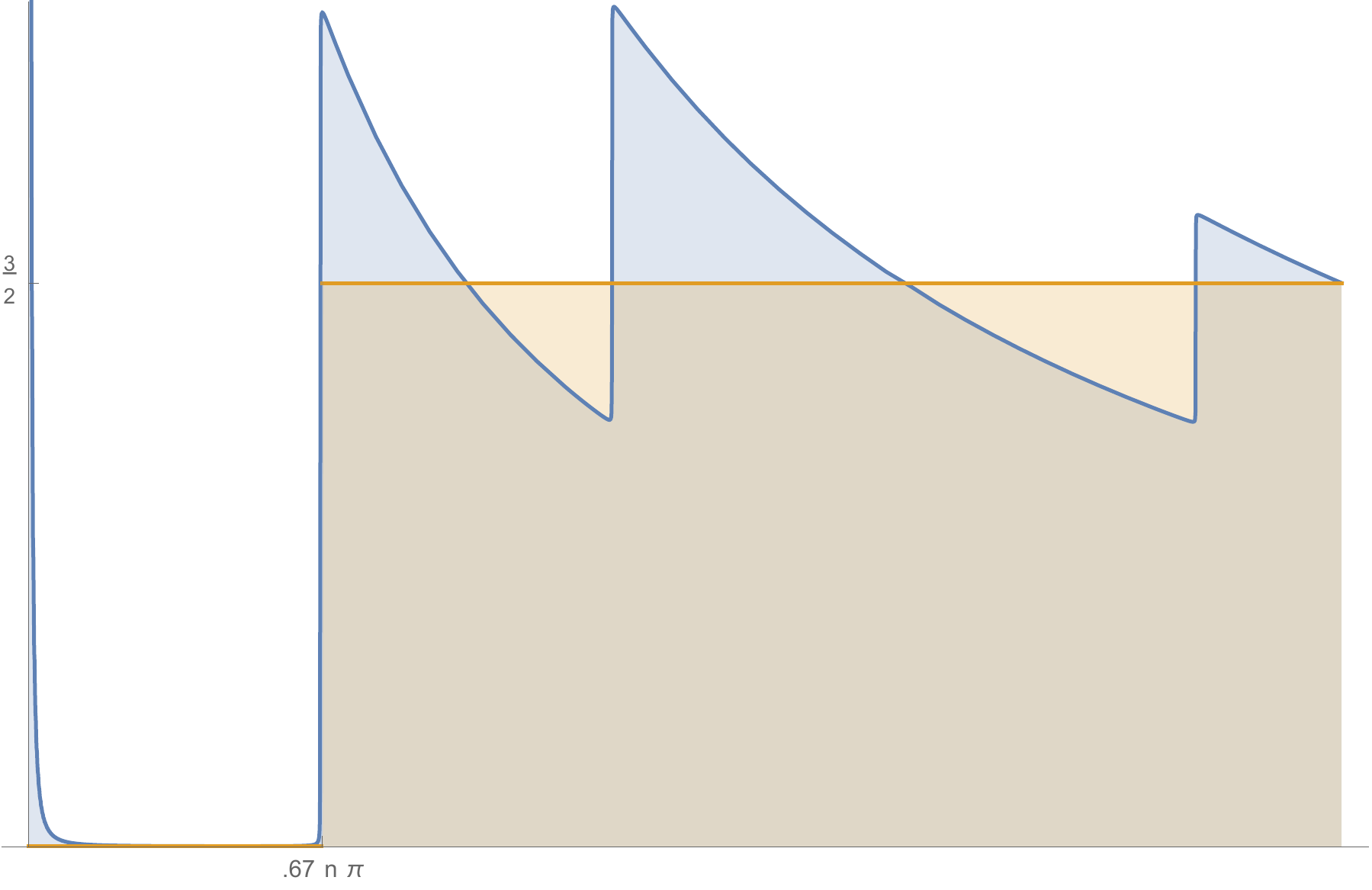}}
\qquad\qquad
{\includegraphics[width=.35\textwidth]{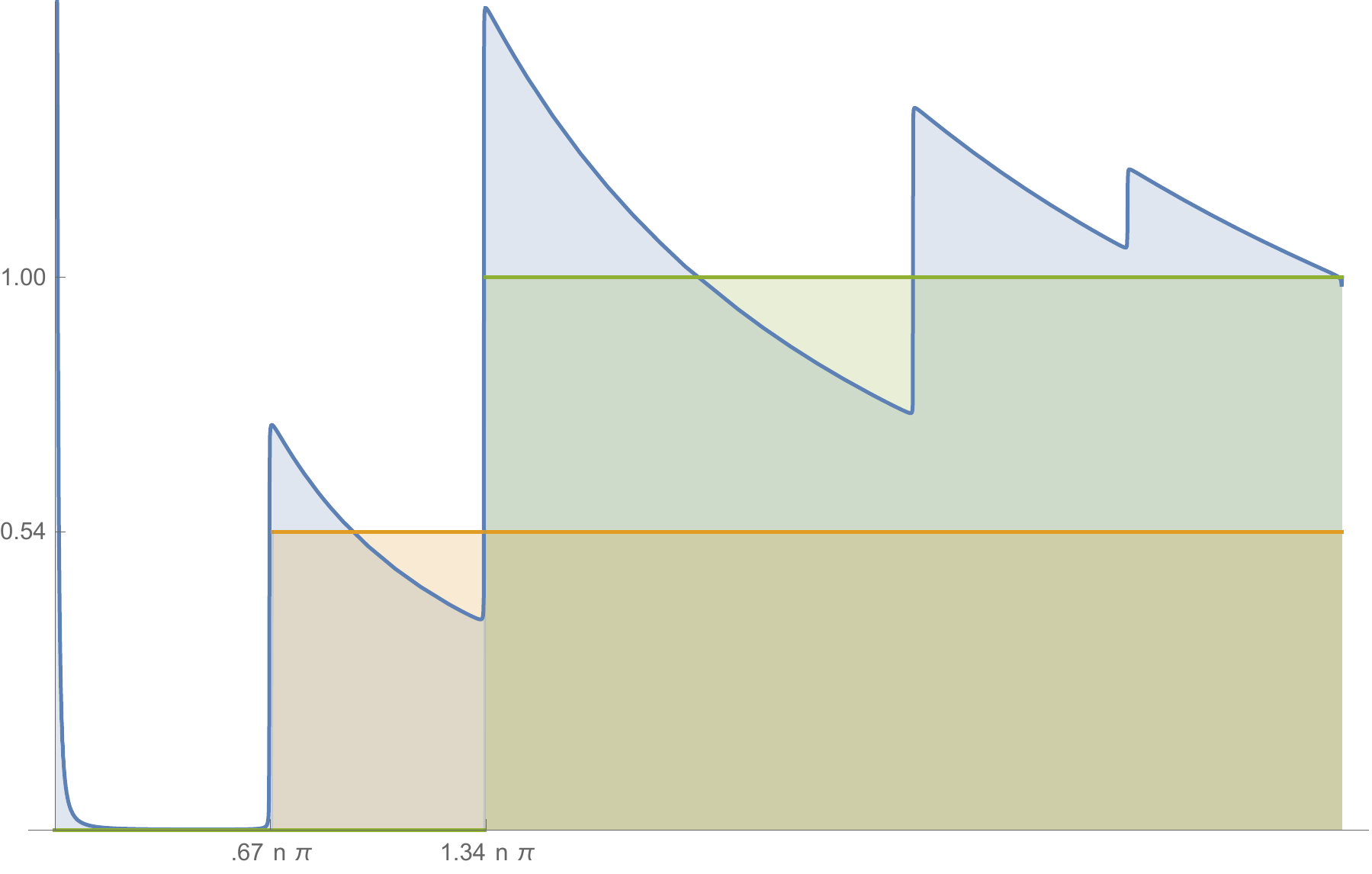}}
\caption{The progressively time-averaged correlator contributions of individual modes $\tilde{C}_n(t)$ for even $n$ (left; $n=100,500,1500$) and odd $n$ (right; $n=101, 501, 1501$). We have also marked the $\Theta$-functions from eq.~(\ref{thetas}).}
\label{cns}
\end{figure}

For all except very small values of $n$, the quantities $\tilde{C}_n(t)$ defined in \eqref{eq:Ctilde} show a universal behavior.\footnote{Most features we discuss are clearly discernible already at $n \gtrsim 10$. We will not need to set the range of $n$ more precisely, because our use of the grand canonical ensemble and approximating sums by integrals are greater sources of error.} Following a rapid decay from their initial values, the $\tilde{C}_n(t)$ hover near zero for a time $\sim 0.67 n \pi$. At that time, the even $n$ quantities undergo one sharp jump to near their asymptotic value of 3/2 and, thereafter, many smaller jumps and gentle decays that keep the $\tilde{C}_n(t)$ near 3/2. For odd $n$, the $\tilde{C}_n(t)$ rise to near their asymptotic value of 1 in two distinct sharp jumps that happen at approximately $0.67 n \pi$ and $1.34 n \pi$, also followed by many smaller jumps and gentle decays which keep the $\tilde{C}_n(t)$ near 1. We have not derived these statements analytically, but they are manifest from the plots in Fig.~\ref{cns}. 

As a coarse approximation to the time dependence of $\tilde{G}(t)$, we may model the $\tilde{C}_n(t)$ as simple step functions:
\begin{equation}
\mathcal{C}_{n~{\rm even}} = \frac{3}{2} \Theta(t - 0.67 n \pi)
\qquad {\rm and} \qquad 
\mathcal{C}_{n~{\rm odd}} = 0.54 \Theta(t - 0.67 n \pi) + 
0.46 \Theta(t - 1.34 n \pi).
\label{thetas}
\end{equation}
In this treatment, the ramp is built up as successive modes shoot up from zero to their final values. This leads to
\begin{equation}
\tilde{G}(t) \approx \frac{1}{N} \left( 
\frac{3}{2} \sum_{n \ {\rm even}}^{t/0.67\pi} N_n +\,
0.54\! \sum_{n \ {\rm odd}}^{t/0.67\pi} N_n +\,
0.46\! \sum_{n \ {\rm odd}}^{t/1.34\pi} N_n
\right).
\label{carefuln}
\end{equation}
As a next step, we substitute the occupation numbers for the typical state and replace the sums with integrals. After these approximations, it will not be meaningful to keep track of the various $\mathcal{O}(1)$ coefficients appearing in (\ref{carefuln}). Thus, we introduce a single $\mathcal{}O(1)$ coefficient $\gamma$ that parameterizes the average rate at which the successive modes join the ramp:
\begin{equation}
\tilde{G}(t) 
\approx 
\frac{1}{N} \cdot \frac{5}{4}\, \sum_{n=1}^{t/\gamma} \frac{8}{\sinh \eta n} 
\approx
\frac{10}{N\eta} \int_{\delta \eta}^{t \eta / \gamma} \frac{du}{\sinh u}
=
\frac{10}{N\eta} \log\frac{\tanh t\eta/2\gamma}{\tanh \delta\eta/2}.
\label{gtcompl}
\end{equation}
The factor of 5/4 is the average height of the jumps undergone by the even (3/2) and odd ($1=0.54 + 0.46$) modes. Since $N = 2\pi^2 / \eta^2 \gg 1$, this reduces to:
\begin{equation}
\tilde{G}(t) 
\approx
\frac{5\eta}{\pi^2} 
\log\left(\frac{2}{\delta\eta}\tanh \frac{t\eta}{2\gamma}\right).
\label{gtsimple}
\end{equation}

\bibliographystyle{utphys}
\bibliography{d1d5chaos}

\end{document}